\documentclass[prd,aps,nofootinbib,showpacs,showkeys]{revtex4}
\def\Xint#1{\mathchoice
   {\XXint\displaystyle\textstyle{#1}}%
   {\XXint\textstyle\scriptstyle{#1}}%
   {\XXint\scriptstyle\scriptscriptstyle{#1}}%
   {\XXint\scriptscriptstyle\scriptscriptstyle{#1}}%
   \!\int}
\def\XXint#1#2#3{{\setbox0=\hbox{$#1{#2#3}{\int}$}
     \vcenter{\hbox{$#2#3$}}\kern-.5\wd0}}

\def\dashint{\Xint-}
\usepackage{graphicx}
\usepackage{bm}
\usepackage{amssymb}
\usepackage{amsmath}
\usepackage{euscript}

\begin{document}

\title{Long-Range Behavior in Quantum Gravity}\thanks{A chapter
contributed to ``Progress in General Relativity and Quantum
Cosmology Research'' (Nova Science Publishers).}

\author{Kirill~A.~Kazakov}\email{kirill@phys.msu.ru}

\affiliation{Department of Theoretical Physics,
Physics Faculty,\\
Moscow State University, $119899$, Moscow, Russian Federation}

\begin{abstract}
Quantum gravity effects of zeroth order in the Planck constant are
investigated in the framework of the low-energy effective theory.
A special emphasis is placed on establishing the correspondence
between classical and quantum theories, for which purpose
transformation properties of the $\hbar^0$-order radiative
contributions to the effective gravitational field under
deformations of a reference frame are determined. Using the
Batalin-Vilkovisky formalism it is shown that the one-loop
contributions violate the principle of general covariance, in the
sense that the quantities which are classically invariant under
such deformations take generally different values in different
reference frames. In particular, variation of the scalar curvature
under transitions between different reference frames is calculated
explicitly. Furthermore, the long-range properties of the
two-point correlation function of the gravitational field are
examined. Using the Schwinger-Keldysh formalism it is proved that
this function is finite in the coincidence limit outside the
region of particle localization. In this limit, the leading term
in the long-range expansion of the correlation function is
calculated explicitly, and the relative value of the root mean
square fluctuation of the Newton potential is found to be
$1/\sqrt{2}.$ It shown also that in the case of a macroscopic
gravitating body, the terms violating general covariance, and the
field fluctuation are both suppressed by a factor $1/N,$ where $N$
is the number of particles in the body. This leads naturally to a
macroscopic formulation of the correspondence between classical
and quantum theories of gravitation. As an application of the
obtained results, the secular precession of a test particle orbit
in the field of a black hole is determined.
\end{abstract}

\maketitle

\section{Introduction}

In this Chapter, quantum properties of the gravitational fields
produced by quantized matter will be considered in the framework
of the low-energy effective theory. This means that our interest
will be in quantum aspects of the gravitational interaction at
energy scales much lower than the Planck energy $$E_{\rm P} =
\sqrt{\frac{\hbar c^5}{G}}\ .$$ From the practical point of view,
this does not present a limitation, since all phenomena observed
in the Universe so far fall perfectly into this category. In view
of the extreme smallness of the Newton constant $G,$ these
phenomena are well described by the lowest-order Einstein theory.
On the other hand, the results obtained within the lowest-order
approximation are model-independent, i.e., they play the role of
low-energy theorems. This universality allows one to draw a number
of important conclusions concerning the synthesis of quantum
theory and gravitation.

An essential part of investigation of this synthesis consists in
establishing a correspondence between classical and quantum
theories. The formal rules of this correspondence are contained in
the Bohr correspondence principle which gives a general recipe for
the construction of operators for various physical quantities.
Informally, the problem of correspondence is in elucidating actual
conditions to be imposed on a system to allow its classical
consideration. Identification of these quasi-classical conditions
constitutes an integral part of interpretative basis of quantum
theory. Right from this point of view the low-energy behavior in
quantum gravity will be considered below.

Two characteristics of the gravitational field produced by a
quantized system will be of our main concern in this Chapter --
the mean value of the gravitational field, and its correlation
function. The results of independent investigation of these
quantities will lead us to one and the same macroscopic
formulation of the quasi-classical conditions in quantum gravity.

Turning to a more detailed formulation of our approach, let us
note, first of all, that the low-energy condition stated above
implies a similar condition on the characteristic length scale $L$
of the process under consideration, $L\gg l_{\rm P},$ where
$l_{\rm P}$ is the Planck length
\begin{eqnarray}\label{lpscale}
l_{\rm P} = \sqrt{\frac{G\hbar}{c^3}}\ .
\end{eqnarray}
\noindent In other words, our interest is in the {\it long-range}
properties of the gravitational field produced by a given system.
As far as this system can be treated classically, $l_{\rm P}$ {\it
is} the length scale characterizing quantum properties of its
gravitational interactions, because $l^2_{\rm P}$ is the only
parameter entering the theory of quantized gravitational field
through the Einstein action
\begin{eqnarray}&&\label{actionh}
S_{\rm g} = - \frac{c^3}{16\pi G}{\displaystyle\int} d^4 x
\sqrt{-g}R\,.
\end{eqnarray}
\noindent In the presence of quantized matter, however, another
parameter with dimension of length comes into play, namely the
gravitational radius
$$r_{\rm g} = \frac{2 G m}{c^2}\,.$$
This parameter appears, of course, already in classical theory.
The question of crucial importance is whether $r_{\rm g}$ has an
independent meaning in quantum domain, representing a scale of
specifically quantum phenomena. This question may seem strange at
first sight, as $r_{\rm g}$ does not contain the Planck constant
$\hbar,$ an inalienable attribute of quantum theory. However, well
known is the fact that gravitational radiative corrections do
contain pieces independent of $\hbar.$ In the framework of the
effective theory, they appear as a power series in $r_{\rm g}/r,$
just like post-Newtonian corrections in classical general
relativity. This fact was first clearly stated by Iwasaki
\cite{iwasaki}. The reason for the appearance of $\hbar^0$ terms
through the loop contributions is that in the case of
gravitational interaction, the mass and ``kinetic'' terms in a
matter Lagrangian determine not only the properties of matter
quanta propagation, but also their couplings. Thus the mass term
of, e.g., scalar field Lagrangian generates the vertices
proportional to
$$\left(\frac{m c}{\hbar}\right)^2$$
containing inverse powers of $\hbar.$ Naively, one expects these
be cancelled by $\hbar$'s coming from the propagators when
combining an amplitude. One should remember, however, that such a
counting of powers of $\hbar$ in Feynman diagrams is a bad
helpmate in the presence of massless particles. Virtual
propagation of gravitons interacting with matter field quanta near
their mass shells results in a {\it root} non-analyticity of the
massive particle form factors at zero momentum transfer ($p$). For
instance, the low-energy expansion of the one-loop diagram in
Fig.~\ref{fig2}(a) begins with terms proportional to
$(-p^2)^{-1/2}\,,$ rather than integer powers (or logarithms) of
$p^2.$ It is this singularity that is responsible for the
appearance of powers of $r_{\rm g}/r$ in the long-range expansion
of the loop contributions. The question we ask is of what nature,
classical or quantum, these pieces are. This is precisely the
question of correspondence in quantum gravity.

Suppose that the matter producing gravitational field satisfies
the usual quantum mechanical quasi-classical conditions, e.g.,
consider sufficiently heavy particles. Then the quasi-classical
conditions for the gravitational field can be inferred from the
requirement that the mean value of the spacetime metric, $\langle
g_{\mu\nu}\rangle,$ coincides with the classical solution of
Einstein equations, corresponding to the same distribution of
gravitating matter. Practically, the most convenient way of
looking for these conditions is to compare {\it transformation
properties} of the quantities involved under deformations of the
reference frame, thus avoiding explicit calculation of the
expectation values. The latter point of view takes advantage of
the fact that the transformation law of classical solutions is
known in advance. Hence, we have to check whether $\langle
g_{\mu\nu}\rangle$ transforms covariantly with respect to
transitions between different reference frames. In other words, we
have to consider the question of {\it general covariance in
quantum gravity.} This is the way we follow in
Sec.~\ref{transformegf}. Alternatively, one can infer the
quasi-classical conditions from the requirement of {\it vanishing
of field fluctuations}. We take this rout in Sec.~\ref{qflucs}.
Some results of the Schwinger-Keldysh and Batalin-Vilkovisky
formalisms, used in our investigation, are summarized in
Sec.~\ref{preliminaries}. Section \ref{conclud} contains
concluding remarks.

Condensed notations of DeWitt \cite{dewitt1} are in force
throughout this chapter. Also, right and left derivatives with
respect to the fields and the sources, respectively, are used. The
dimensional regularization of all divergent quantities is assumed.

\section{Preliminaries}\label{preliminaries}

Before going into detailed discussion of the question of general
covariance in quantum gravity, let us describe the general setting
we will be working in.

\subsection{Frame of reference and interacting fields.}

First of all, we should set a frame of reference, {\it i.e.}, a
system of idealized reference bodies with respect to which the
4-position in spacetime can be fixed. Let us assume, for
definiteness, that the frame of reference is realized by means of
an appropriate distribution of electrically charged matter. For
simplicity, the energy-momentum of matter, as well as of the
electromagnetic field it produces, will be assumed sufficiently
small so as not to alter the gravitational field under
consideration. The 4-position in spacetime can be determined by
exchanging electromagnetic signals with a number of charged matter
species. The electric charge distributions $\sigma_a$ of the
latter are thus supposed to be in a one-to-one correspondence with
the spacetime coordinates $x_{\mu},$ $$\sigma_a\leftrightarrow
x_{\mu}\,,$$ where index $a$ enumerates the species. The
$\sigma_a(x)$ will be assumed smooth scalar functions. Physical
properties of the reference frame are determined by the action
$S_{\sigma}$ which specific form is of no importance for us.

Next, let us consider a system of interacting gravitational and
matter fields. The latter are arbitrary species, bosons or
fermions, self-interacting or not, denoted collectively by
$\phi_i,$ $i = 1,2,...,k.$ Dynamical variables of the
gravitational field are $h_{\mu\nu} = g_{\mu\nu} - \eta_{\mu\nu}.$
Dynamics of the system is described by the action $S_{\rm g} +
S_{\phi},$ where $S_{\phi}$ is the matter action, and $S_{\rm g}$
is given\footnote{Our notation is $R_{\mu\nu} \equiv
R^{\alpha}_{~\mu\alpha\nu} =
\partial_{\alpha}\Gamma^{\alpha}_{\mu\nu} - \cdot\cdot\cdot,
~R \equiv R_{\mu\nu} g^{\mu\nu}, ~g\equiv \det g_{\mu\nu},
~g_{\mu\nu} = {\rm sgn}(+,-,-,-),$ $\eta_{\mu\nu} = {\rm
diag}\{+1,-1,-1,-1\}.$ The Minkowski tensor $\eta$ is used to
raise and lower tensor indices. The units in which $\hbar = c =
16\pi G = 1$ are chosen in what follows.} by Eq.~(\ref{actionh}).

The total action $S = S_{\rm g} + S_{\phi} + S_{\sigma}$ is
invariant under the gauge transformations
\begin{eqnarray}\label{gaugesymm}
\delta h_{\mu\nu} &=& \xi^{\alpha}\partial_{\alpha}h_{\mu\nu} +
(\eta_{\mu\alpha} + h_{\mu\alpha})\partial_{\nu}\xi^{\alpha} +
(\eta_{\nu\alpha} + h_{\nu\alpha})\partial_{\mu}\xi^{\alpha}
\equiv G_{\mu\nu}^{\alpha}\xi_{\alpha}\,,
\label{gaugesymmh}\\
\delta\phi_i &=&  G_i^{\alpha}\xi_{\alpha}\,,
\label{gaugesymmphi}\\
\delta\sigma_a &=& \sigma_{a,\alpha}\xi^{\alpha}
\label{gaugesymmsigma}
\end{eqnarray}
\noindent The generators $G_{\mu\nu}, G_i$ span the closed algebra
\begin{eqnarray}\label{algebra}
G_{\mu\nu}^{\alpha,\sigma\lambda} G_{\sigma\lambda}^{\beta} -
G_{\mu\nu}^{\beta,\sigma\lambda} G_{\sigma\lambda}^{\alpha} &=&
f_{~~~\gamma}^{\alpha\beta} G_{\mu\nu}^{\gamma}\,,
\nonumber\\
 G^{\alpha,k}_i  G^{\beta}_k
-  G^{\beta,k}_i  G^{\alpha}_k &=& f^{\alpha\beta}_{~~~\gamma}
G^{\gamma}_i\,,
\end{eqnarray}
where the "structure constants" $f^{\alpha\beta}_{~~~\gamma}$ are
defined by
\begin{eqnarray}&&
f_{~~~\gamma}^{\alpha\beta}\xi_{\alpha}\eta_{\beta} =
\xi_{\alpha}\partial^{\alpha}\eta_{\gamma} -
\eta_{\alpha}\partial^{\alpha}\xi_{\gamma}\,.
\end{eqnarray}
\noindent Let the gauge-fixing action be written in the form
\begin{eqnarray}\label{gaugefixpi}&&
S_{\rm gf} = \left(F_{\alpha} -
\frac{1}{2}\pi^{\beta}\zeta_{\beta\alpha}\right)\pi^{\alpha}\,,
\end{eqnarray}
\noindent where $F_{\alpha}$ is a set of functions of the fields
$h_{\mu\nu},$ fixing general invariance, $\pi^{\alpha}$ auxiliary
fields introducing the gauge, and $\zeta_{\alpha\beta}$ a
non-degenerate symmetric matrix weighting the functions
$F_{\alpha};$ the particular choice $\zeta_{\alpha\beta} =
\xi\eta_{\alpha\beta}$ corresponds to the well-known Feynman
weighting of the gauge conditions. Introducing the ghost fields
$c_{\alpha},$ $\bar{c}^{\alpha},$ and denoting all the fields
collectively by $\Phi,$ we write the Faddeev-Popov action
\cite{faddeev}
\begin{eqnarray}\label{fp}
S_{\rm FP}[\Phi] = S + S_{\rm gf} +
\bar{c}^{\beta}F_{\beta}^{,\mu\nu}G_{\mu\nu}^{\alpha}c_{\alpha}\,.
\end{eqnarray}
\noindent $S_{\rm FP}[\Phi]$ is invariant under the following
Becchi-Rouet-Stora-Tyutin (BRST) transformations \cite{brst}
\begin{eqnarray}\label{brst}
\delta h_{\mu\nu} &=& G_{\mu\nu}^{\alpha}c_{\alpha}\lambda\,,
\nonumber\\
\delta \phi_i &=&  G_i^{\alpha}c_{\alpha}\lambda\,,
\nonumber\\
\delta \sigma_a &=& \sigma_{a,\alpha}c^{\alpha}\lambda\,,
\nonumber\\
\delta c_{\gamma} &=& - \frac{1}{2}f^{\alpha\beta}_{~~~\gamma}
c_{\alpha}c_{\beta}\lambda\,,
\nonumber\\
\delta \bar{c}^{\alpha} &=& \pi^{\alpha}\lambda\,,
\nonumber\\
\delta\pi^{\alpha} &=& 0\,,
\end{eqnarray}
\noindent where $\lambda$ is a constant anticommuting parameter.
Finally, the generating functional of Green functions has the form
\begin{eqnarray}\label{gener}&&
Z[J,K] = {\displaystyle\int}\EuScript{D}\Phi \exp\{i
(\Sigma[\Phi,K] + \bar{\beta}^{\alpha}c_{\alpha} +
\beta_{\alpha}\bar{c}^{\alpha} + t^{\mu\nu}h_{\mu\nu} + j^i\phi_i
+ s^a\sigma_a)\},
\end{eqnarray}
\noindent where
\begin{eqnarray}&&
\Sigma[\Phi,K] = S_{\rm FP}[\Phi] +
k^{\mu\nu}G_{\mu\nu}^{\alpha}c_{\alpha} + q^i
G_i^{\alpha}c_{\alpha} + r^a\sigma_{a,\alpha}c^{\alpha} -
\frac{l^{\gamma}}{2}
f^{\alpha\beta}_{~~~\gamma}c_{\alpha}c_{\beta} +
n_{\alpha}\pi^{\alpha}\,, \nonumber
\end{eqnarray}\noindent
\{$t,$ $j,$ $s,$ $\bar{\beta},$ $\beta,$ $0$\} $\equiv J$ ordinary
sources, and \{$k,$ $q,$ $r,$ $l,$ $n,$ $0$\} $\equiv K$ the
BRST-transformation sources \cite{zinnjustin} for the fields
\{$h,$ $\phi,$ $\sigma,$ $c,$ $\bar{c},$ $\pi$\} $\equiv \Phi,$
respectively, and integration is carried over all field
configurations satisfying
$$\Phi^{\pm}\to 0\,, \qquad {\rm for}\quad t \to
\mp\infty\,,$$ where the superscripts $+$ and $-$ denote the
positive- and negative-frequency parts of the fields,
respectively. The $K$-sources are introduced into $Z$ for the
purpose of studying gauge dependence of observable quantities.

\subsection{Effective fields and correlation functions}

The central objects of our investigation in the sections below
will be the expectation value of the gravitational field produced
by an elementary particle, and its correlation function. They can
be obtained from the one- and two-point Green functions of the
gravitational field, respectively,
\begin{eqnarray}\label{green1}
h^{\rm eff}_{\mu\nu}(j) &=& \left.\frac{\delta W[J,K]}{\delta
t^{\mu\nu}}\right|_{\genfrac{}{}{0pt}{}{J\setminus j = 0}{K =
0}}\,, \\ C_{\mu\nu\alpha\beta}(j) &=& \left.\frac{\delta^2
W[J,K]}{\delta t^{\mu\nu}\delta t^{\alpha\beta}
}\right|_{\genfrac{}{}{0pt}{}{J\setminus j = 0}{K = 0}}\,,
\label{green2}
\end{eqnarray}
\noindent with the help of the well-known reduction formula
applied to the external matter lines to go over onto the mass
shell of the particle. In Eqs.~(\ref{green1}), (\ref{green2}), $W$
is the generating functional of connected Green functions,
$$W[J,K] = - i \ln Z[J,K],$$ and the symbol $J\setminus j$ means
that the source $j$ is excluded from the set $J,$

However, under the Feynman asymptotic conditions specified above,
Eqs.~(\ref{green1}), (\ref{green2}) give the {\it in-out} matrix
elements of operators, rather than the {\it in-in} expectation
values we are interested in. As is well known, in order to find
the latter, the usual Feynman rules for constructing the matrix
elements must be modified. According to the so-called closed time
path formalism of Schwinger and Keldysh \cite{keldysh,schwinger2}
(modern reviews of this formalism can be found in
Refs.~\cite{jordan,paz}), this amounts to using in
Eqs.~(\ref{green1}), (\ref{green2}) the generating functional
\begin{eqnarray}\label{genercpt}
Z_{\rm SK}[J_\pm,K_\pm] =
\int\EuScript{D}\Phi_-\int\EuScript{D}\Phi_+ \exp\{i
(\Sigma[\Phi_{\pm},K_{\pm}] &+& \bar{\beta}^{\alpha}c_{\alpha} +
\beta_{\alpha}\bar{c}^{\alpha} + t^{\mu\nu}h_{\mu\nu} \nonumber\\
&+& j^i\phi_i + s^a\sigma_a)\},
\end{eqnarray}
\noindent where
\begin{eqnarray}&&
\Sigma[\Phi_{\pm},K_{\pm}] = S_{\rm FP}[\Phi_+] - S_{\rm
FP}[\Phi_-] + k^{\mu\nu}G_{\mu\nu}^{\alpha}c_{\alpha} + q^i
G_i^{\alpha}c_{\alpha} + r^a\sigma_{a,\alpha}c^{\alpha} -
\frac{l^{\gamma}}{2}
f^{\alpha\beta}_{~~~\gamma}c_{\alpha}c_{\beta} +
n_{\alpha}\pi^{\alpha}\,, \nonumber
\end{eqnarray}
instead of (\ref{gener}). Here the subscript $+$ ($-$) shows that
the time argument of the integration variable runs from $-\infty$
to $+\infty$ (from $+\infty$ to $-\infty$). Integration is over
all fields satisfying
\begin{eqnarray}\label{bcond}
\Phi^{+} \to 0\,, \qquad {\rm for} \quad t &\to& -\infty\,,
\nonumber\\ \Phi_{+} = \Phi_{-} \qquad {\rm for} \quad t &\to& +
\infty\,.
\end{eqnarray}\noindent We do not distinguish the plus and minus
field components in the source terms explicitly, implying that
summation over repeated Greek indices includes summation over
$\pm$ as well as spacetime integration. For instance,
$$t^{\mu\nu}h_{\mu\nu} \equiv \int d^4 x \left(t_+^{\mu\nu}h_{\mu\nu+} +
t_-^{\mu\nu}h_{\mu\nu-}\right)\,,$$ $$l^{\gamma}
f^{\alpha\beta}_{~~~\gamma}c_{\alpha}c_{\beta} \equiv \int d^4 x
\left(l_+^{\gamma}
f^{\alpha\beta}_{~~~\gamma}c_{\alpha+}c_{\beta+} + l_-^{\gamma}
f^{\alpha\beta}_{~~~\gamma}c_{\alpha-}c_{\beta-}\right)\,, \qquad
{\rm etc.}
$$ With the help of the new generating functional, the in-in
expectation value of, e.g., the product
$\hat{h}_{\mu\nu}(x)\hat{h}_{\alpha\beta}(x')$ can be written as
\begin{eqnarray}\label{fintctp1} \langle {\rm in}
|\hat{h}_{\mu\nu}(x)\hat{h}_{\alpha\beta}(x')|{\rm in}\rangle &=&
-Z^{-1}_{\rm SK}\left.\frac{\delta^2 Z_{\rm
SK}[J_{\pm},K_{\pm}]}{\delta t_-^{\mu\nu}(x)\delta
t_+^{\alpha\beta}(x')}\right|_{\genfrac{}{}{0pt}{}{J\setminus j =
0}{K = 0}}\,.
\end{eqnarray}
\noindent It is seen that $\langle {\rm in}
|\hat{h}^{\mu\nu}(x)\hat{h}^{\alpha\beta}(x')|{\rm in}\rangle$ is
given by the ordinary functional integral but with the number of
fields doubled, and unusual boundary conditions specified above.
Accordingly, diagrammatics generated upon expanding this integral
in perturbation theory consists of the following elements. There
are four types of pairings for each component of the set $\Phi,$
corresponding to the four different ways of placing two field
operators on the two branches of the time path. They are
conveniently combined into $2 \times 2$ matrices according
to\footnote{Gothic letters are used to distinguish quantities
representing columns, matrices etc. with respect to indices
$+,-.$}
$$\mathfrak{D} = \left(
\begin{array}{cc}
D_{++}&D_{+-}\\
D_{-+}&D_{--}
\end{array}\right)\,.$$
For instance, pairings matrices for the gravitational and scalar
matter fields read
$$\mathfrak{D}_{ik}(x,y) = \left(\begin{array}{cc}i\langle
T\hat{\phi}_i(x)\hat{\phi}_k(y)\rangle_0 & i\langle
\hat{\phi}_k(y)\hat{\phi}_i(x)\rangle_0\\i\langle
\hat{\phi}_i(x)\hat{\phi}_k(y)\rangle_0 & i\langle
\tilde{T}\hat{\phi}_i(x)\hat{\phi}_k(y)\rangle_0\end{array}\right)\,,
$$
$$\mathfrak{D}_{\mu\nu\alpha\beta}(x,y) =
\left(\begin{array}{cc}i\langle
T\hat{h}_{\mu\nu}(x)\hat{h}_{\alpha\beta}(y)\rangle_0 & i\langle
\hat{h}_{\alpha\beta}(y)\hat{h}_{\mu\nu}(x)\rangle_0\\
i\langle \hat{h}_{\mu\nu}(x)\hat{h}_{\alpha\beta}(y)\rangle_0 &
i\langle
\tilde{T}\hat{h}_{\mu\nu}(x)\hat{h}_{\alpha\beta}(y)\rangle_0\end{array}\right)\,,
$$
where the operation of time ordering $T$ ($\tilde{T}$) arranges
the factors so that the time arguments decrease (increase) from
left to right, and $\langle\cdot\rangle_0$ denotes vacuum
averaging. The ``propagators'' $\mathfrak{D}_{\mu\nu\alpha\beta},$
$\mathfrak{D}_{ik}$ satisfy the following matrix equations
\begin{eqnarray}\label{aprop}
\mathfrak{G}^{\mu\nu\alpha\beta}
\mathfrak{D}_{\alpha\beta\sigma\lambda} &=& -
\mathfrak{e}\delta^{\mu\nu}_{\sigma\lambda}\,,
\quad\mathfrak{G}^{\mu\nu\alpha\beta} =
\mathfrak{i}~\frac{\delta^2 S^{(2)}}{\delta
h_{\mu\nu}\delta h_{\alpha\beta}}\,, \quad
\delta^{\mu\nu}_{\sigma\lambda} = \frac{1}{2}
\left(\delta_{\sigma}^{\mu}\delta_{\lambda}^{\nu}
+ \delta_{\sigma}^{\nu}\delta_{\lambda}^{\mu}\right)\\
\mathfrak{G}^{ik}\mathfrak{D}_{kl} &=& -
\mathfrak{e}\delta^{i}_{l}\,, \quad\mathfrak{G}^{ik} =
\mathfrak{i}~\frac{\delta^2S^{(2)}}{\delta\phi_i\delta
\phi_k}\,,\label{aprop1}
\end{eqnarray}
\noindent where $S^{(2)}$ denotes the free field part of the
gauge-fixed action after the gauge introducing fields
$\pi^{\alpha}$ have been integrated out, and $\mathfrak{e},$
$\mathfrak{i}$ are $2\times2$ matrices with respect to indices
$+,-:$
$$\mathfrak{e} = \left(\begin{array}{cc}
1&0\\0&1\end{array}\right)\,, \quad \mathfrak{i} =
\left(\begin{array}{cc} 1&0\\0&-1\end{array}\right)\,.$$ As in the
ordinary Feynman diagrammatics of the S-matrix theory, the
propagators are contracted with the vertex factors generated by
the interaction part of the action, $S^{\rm int}[\Phi],$ with
subsequent summation over $(+,-)$ in the vertices, each ``$-$''
vertex coming with an extra factor $(-1).$ This can be represented
as the matrix multiplication of
$\mathfrak{D}_{\mu\nu\alpha\beta},\mathfrak{D}_{ik}$ with suitable
matrix vertices. For instance, the $\phi^2 h$ part of the action
generates the matrix vertex $\mathfrak{V}^{\mu\nu,ik}$ which in
components has the form
$$V_{stu}^{\mu\nu,ik}(x,y,z) =
e_{stu}\left.\frac{\delta^3 S}{\delta h_{\mu\nu}(x)
\delta\phi_i(y)\delta\phi_k(z)}\right|_{\Phi = 0}\,,$$ where
indices $s,t,u$ take the values $+,-,$ and $e_{stu}$ is defined by
$e_{+++} = - e_{---} = 1$ and zero otherwise. An external $\phi$
line is represented in this notation by a column
$$\mathfrak{r}_i = \left(\begin{array}{c} \bar{\phi}_i \\ \bar{\phi}_i
\end{array}\right),$$ satisfying
\begin{eqnarray}\label{free}
\mathfrak{G}^{ik}\mathfrak{r}_k = \left(\begin{array}{c} 0 \\
0 \end{array}\right)\,.
\end{eqnarray}
\noindent

For future references, we give explicit expressions for various
pairings of a single real scalar field
\begin{eqnarray}\label{explprop}
D_{++}(x,y) &=& \int\frac{d^4 k}{(2\pi)^4}\frac{e^{-ik(x-y)}}{m^2
- k^2 - i0}\,, \quad D_{--}(x,y) =
\int\frac{d^4 k}{(2\pi)^4}\frac{e^{-ik(x-y)}}{k^2 - m^2 - i0}\,,\nonumber\\
D_{-+}(x,y) &=& i\int\frac{d^4 k}{(2\pi)^3}\theta(k^0)\delta(k^2 -
m^2)e^{-ik(x-y)}\,, \quad D_{+-}(x,y) =  D_{-+}(y,x)\,.
\end{eqnarray}
\noindent

\subsection{Generating functionals and Slavnov identities.}

The functional $\Sigma[\Phi_{\pm},K_{\pm}]$ can be written as
\cite{batvil}
$$\Sigma[\Phi_{\pm},K_{\pm}] =
\Sigma^r\left[\Phi_{\pm},\frac{\delta\Psi[\Phi_{\pm},K_{\pm}]}{\delta\Phi_{\pm}}\right]\,,
$$ where the so-called reduced action $\Sigma^r$ and the gauge
fermion $\Psi$ are defined by
\begin{eqnarray}&&
\Sigma^r[\Phi_{\pm},K_{\pm}] = S +
k^{\mu\nu}G_{\mu\nu}^{\alpha}c_{\alpha} + q^i
G_i^{\alpha}c_{\alpha} + r^a\sigma_{a,\alpha}c^{\alpha} -
\frac{l^{\gamma}}{2}
f^{\alpha\beta}_{~~~\gamma}c_{\alpha}c_{\beta} +
n_{\alpha}\pi^{\alpha}\,, \nonumber
\end{eqnarray}
\noindent $$\Psi[\Phi_{\pm},K_{\pm}] = k^{\mu\nu}h_{\mu\nu} +
q^i\phi_i + r^a\sigma_a + l^{\alpha}c_{\alpha} +
n_{\alpha}\bar{c}^{\alpha} + \bar{c}^{\alpha}\left(F_{\alpha} -
\frac{1}{2}\pi^{\beta}\zeta_{\beta\alpha}\right).$$ Let us further
simplify notation abbreviating $k^{\mu\nu}h_{\mu\nu} + q^i\phi_i +
r^a\sigma_a + l^{\alpha}c_{\alpha} + n_{\alpha}\bar{c}^{\alpha}
\equiv K\Phi,$ and similarly for other sums of products of fields
and sources or derivatives with respect to $\Phi, J, K.$ Omitting
also the subscript $\pm,$ the variation of $\Sigma$ under
infinitesimal variation of the gauge conditions takes the form
\begin{eqnarray}&&
\delta\Sigma[\Phi,K] = \frac{\delta\Delta\Psi[\Phi,K]}{\delta\Phi}
\frac{\delta\Sigma[\Phi,K]}{\delta K}\,. \nonumber
\end{eqnarray}
\noindent The corresponding variation of the generating functional
\begin{eqnarray}&&
\delta Z_{\rm SK}[J,K] = i
\int\EuScript{D}\Phi_-\int\EuScript{D}\Phi_+ \exp\{i (\Sigma +
J\Phi)\} \frac{\delta\Delta\Psi(\Phi,K)}{\delta\Phi}
\frac{\delta\Sigma(\Phi,K)}{\delta K}\,. \nonumber
\end{eqnarray}
\noindent Integrating by parts and omitting $\delta^2
\Sigma/\delta K\delta\Phi\sim\delta^{(4)}(0)$ in the latter
equation gives
\begin{eqnarray}&&\label{zvar}
\delta Z_{\rm SK}[J,K] = i J\frac{\delta}{\delta
K}\int\EuScript{D}\Phi_-\int\EuScript{D}\Phi_+ \Delta\Psi\exp\{i
(\Sigma + J\Phi)\}\,.
\end{eqnarray}
\noindent Since $\Sigma$ is invariant under the BRST
transformation (\ref{brst}), a BRST change of integration
variables in Eq.~(\ref{genercpt}) gives the Slavnov identity for
the generating functional
\begin{eqnarray}\label{slav}
J\frac{\delta Z_{\rm SK}}{\delta K} = 0\,,
\end{eqnarray}
\noindent which allows one to rewrite Eq.~(\ref{zvar}) in terms of
the generating functional of connected Green functions as
\begin{eqnarray}&&\label{wvar}
\delta W_{\rm SK}[J,K] =  J\frac{\delta}{\delta K}\langle
\Delta\Psi\rangle \,,
\end{eqnarray}
\noindent where $\langle X\rangle$ denotes the functional
averaging of $X$ \cite{tyutin}.

\section{Transformation properties of effective gravitational
field}\label{transformegf}

In this section, we will investigate the structure of the
$\hbar^0$-order contributions to the effective gravitational
field, concentrating mainly on their transformation properties
under deformations of the reference frame. The appearance of such
contributions through the radiative corrections has its roots in
the field properties of quantized matter sources. To make the
field-theoretic aspect of the problem clearer, it is convenient to
get rid of the purely quantum mechanical issues related to the
quantum particle kinematics assuming matter quanta sufficiently
heavy to neglect the position-velocity indeterminacy following
from the Heisenberg principle. This assumption justifies the use
of the term ``particle at rest,'' and allows one to introduce a
fixed distance $r$ between the particle and the point of
observation.

The mean gravitational field produced by a massive particle is a
function of five dimensional parameters -- the fundamental
constants $\hbar,G,c,$ the particle's mass $m,$ and the distance
$r.$ Of these only two independent dimensionless combinations can
be constructed, which we choose to be $\chi = l_{\rm P}/r$ and
$\varkappa = l_{\rm C}/r,$ where $l_{\rm C} = \hbar/mc$ is the
Compton length of the particle. The above assumption means that
the gravitational field is considered in the limit $\varkappa \to
0.$ For a fixed particle's mass, small values of $\varkappa$ imply
large values of $r,$ leading naturally to the long-range expansion
of the mean gravitational field. Accordingly, we will use where
appropriate the terminology and general ideas of effective field
theories \cite{weinberg1,weinberg2,donoghue1,donoghue}.

\subsection{General covariance at the tree level.}\label{covtree}

\begin{figure}
\scalebox{0.8}{\includegraphics{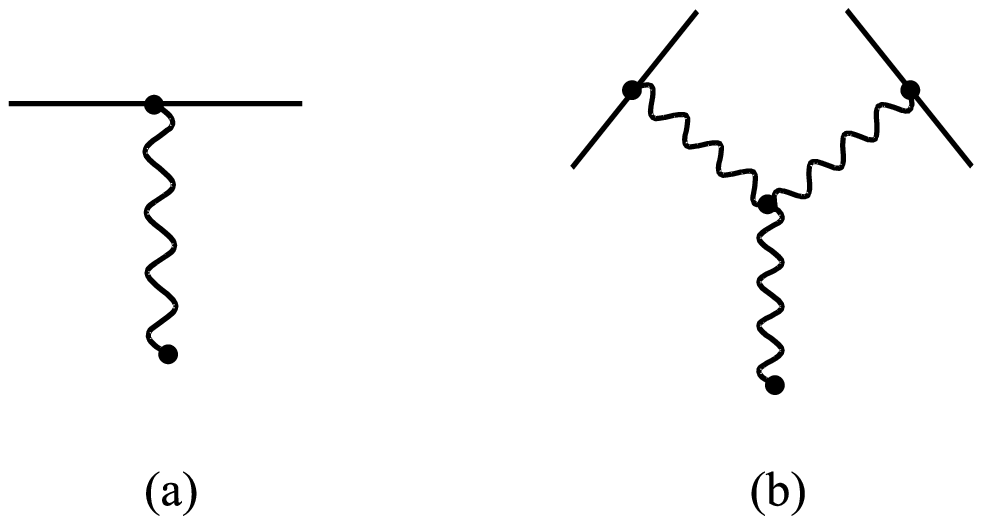}} \caption{Lower-order
tree diagrams representing solution of the Einstein equations. (a)
The first order (Newtonian) term. (b) One of the second-order
diagrams describing the first post-Newtonian correction. Wavy
lines represent gravitons, solid lines the auxiliary source
$t^{\mu\nu},$ or expectation value of the energy-momentum tensor
of matter, $\langle \hat{T}^{\mu\nu} \rangle.$} \label{fig1}
\end{figure}

From the point of view of the general formalism outlined in the
preceding sections, the classical Einstein theory corresponds to
the tree approximation of the full quantum theory. The tree
contributions to the expectation values of field operators
coincide with the corresponding classical fields, and the
effective equations of motion
$$\left\langle \frac{\delta\Sigma}{\delta h_{\mu\nu}} \right\rangle
+ t^{\mu\nu} = 0\,,$$ expressing the translation invariance of the
functional integral measure, go over into the classical Einstein
equations. Some of the tree diagrams representing their solution
are shown in Fig.~\ref{fig1}. The results of the preceding section
allow one, in particular, to reestablish the general covariance of
these equations.

As was mentioned in the Introduction, coordinate transformations
are replaced in the quantum theory by the field transformations.
In particular, transformations of the reference frame are
represented by variations of the fields $\sigma_i,$ induced by
appropriate variations of the gauge conditions. Namely, it follows
from Eq.~(\ref{wvar}) at the tree level that a gauge variation
$\Delta F_{\alpha}$ induces the following variations of the metric
and reference fields
\begin{eqnarray}\label{vartreeg}
\delta g_{\mu\nu} &=& g_{\mu\nu,\alpha}\Xi^{\alpha} +
g_{\mu\alpha}\Xi^{\alpha}_{,\nu}
+ g_{\nu\alpha}\Xi^{\alpha}_{,\mu} \,, \\
\delta\sigma_a &=& \sigma_{a,\alpha}\Xi^{\alpha}\,, \qquad
\Xi_{\alpha} = \langle c_{\alpha}\Delta\Psi\rangle\,,
\label{vartreesigma}
\end{eqnarray}
\noindent where $\Delta\Psi$ is the corresponding variation of the
gauge fermion.

The functions $g_{\mu\nu},\sigma_a$ undergo the same variations
(\ref{vartreeg}), (\ref{vartreesigma}) under the spacetime
diffeomorphism
$$x^{\mu} \to x^{\mu} + \delta x^{\mu},$$
with $\delta x^{\mu} = - \Xi^{\mu}\,.$ Let us consider any
quantity entering the Einstein equations, for instance, the scalar
curvature $R.$ Under the above change of gauge conditions, the
tree value of $R$ measured at a point $\sigma^0$ of the reference
frame remains unchanged,
\begin{eqnarray}\label{example}
\delta R[g(x(\sigma^0))] = \frac{\delta R}{\delta
g_{\mu\nu}}\delta g_{\mu\nu} + \frac{\partial R}{\partial
x^{\mu}}\delta x^{\mu} = \frac{\partial R}{\partial
x^{\alpha}}\Xi^{\alpha} - \frac{\partial R}{\partial
x^{\mu}}\Xi^{\mu} = 0\,.
\end{eqnarray}
\noindent Analogously, the tree contribution to any tensor
quantity $O_{\mu\nu...}$ (or $O^{\mu\nu...}$), for instance, the
metric $g_{\mu\nu}$ itself, calculated at a fixed reference point
$\sigma^0,$ transforms covariantly (contravariantly), as
prescribed by the position of the tensor indices of the
corresponding operator. This is the manifestation of general
covariance of the classical Einstein theory in terms of quantum
field theory.

\subsection{General covariance at the one-loop order.}

Let us now consider the transformation properties of the one-loop
contributions. As in Sec.~\ref{covtree}, we have to determine the
effect of an arbitrary gauge variation on the value of the
effective metric, and also on the functions $\sigma_a,$ {\it
i.e.}, on the structure of the reference frame. After that, the
transformation law of observables, defined generally as the
diffeomorphism-invariant functions of the metric and reference
fields, can be determined in the way followed in
Sec.~\ref{covtree}. For definiteness, we will deal below with the
scalar curvature $R.$ Since we are interested in the one-loop
contribution to the first post-Newtonian correction, we can
linearize $R$ in $h_{\mu\nu}:$
\begin{eqnarray}\label{r}
R = \partial^{\mu}\partial^{\nu} h_{\mu\nu} - \Box h\,, \qquad
h\equiv \eta^{\mu\nu}h_{\mu\nu}\,.
\end{eqnarray}
\noindent

The transformation properties of $R$ under variations of the
weighting matrix $\zeta_{\alpha\beta}$ are considered in
Sec.~\ref{feynman}, and under variations of the gauge functions
$F_{\alpha}$ themselves in Sec.~\ref{general}.

\subsubsection{Dependence of effective metric on weighting parameters.}
\label{feynman}

Dependence of the effective fields on the weighting matrix
$\zeta_{\alpha\beta}$ can be determined in a quite general way
without specifying either the gauge functions $F_{\alpha},$ or the
properties of gravitating matter fields. We begin with the
classical theory in Sec.~\ref{tree}, and then consider the
one-loop order in Sec.~\ref{oneloop}.

\paragraph{The tree level.}\label{tree}

As we saw in Sec.~\ref{covtree}, arbitrary gauge variations lead
to the transformations of the classical fields, equivalent to the
spacetime diffeomorphisms, and thus do not affect the values of
$R$. In the particular case of variations of the weighting matrix,
however, not only $R,$ but also the effective fields themselves
remain unchanged. This means that the structure of reference
frames in classical theory is determined by the functions
$F_{\alpha}$ only. As this differs in the full quantum theory, a
somewhat more detailed discussion of this issue will be given in
this Section.

To demonstrate $\zeta_{\alpha\beta}$-independence of the classical
metric, let us first integrate the auxiliary fields $\pi^{\alpha}$
out of the gauge-fixing action (\ref{gaugefixpi}),
\begin{eqnarray}\label{gaugefixp}&&
S_{\rm gf} \to S^{\xi}_{\rm gf} = \frac{1}{2}
F_{\alpha}\xi^{\alpha\beta} F_{\beta}\,, \qquad
\zeta_{\alpha\beta}\xi^{\beta\gamma} = \delta_{\alpha}^{\gamma}\,.
\end{eqnarray}
\noindent The classical equations of motion thus become
\begin{eqnarray}\label{classeq}&&
\frac{\delta (S_{\rm g} + S^{\xi}_{\rm gf})}{\delta g_{\mu\nu}} =
- T^{\mu\nu},
\end{eqnarray}
\noindent where $T^{\mu\nu}$ is the energy-momentum tensor of
matter. Using invariance of the action $S$ under the gauge
transformations
\begin{eqnarray}\label{gaugesym}
\delta g_{\mu\nu} = \xi^{\alpha}\partial_{\alpha}g_{\mu\nu} +
g_{\mu\alpha}\partial_{\nu}\xi^{\alpha} +
g_{\nu\alpha}\partial_{\mu}\xi^{\alpha} \equiv
\nabla_{\mu}\xi_{\nu}\,,
\end{eqnarray}
\noindent and taking into account the ``conservation law''
$\nabla_{\mu}T^{\mu\nu} = 0,$ one has from Eq.~(\ref{classeq})
\begin{eqnarray}&&\label{cderiv}
F_{\alpha}\xi^{\alpha\beta} \frac{\delta F_{\beta}}{\delta
g_{\mu\nu}(x)} \nabla^{x}_{\mu}\delta(x - y) = 0\,.
\end{eqnarray}
\noindent The matrix $M_{\beta}^{\nu}(x,y) = \delta
F_{\beta}/\delta g_{\mu\nu}(x)\nabla^{x}_{\mu}\delta(x - y)$ is
non-degenerate; its determinant $\Delta \equiv \det
M_{\beta}^{\nu}(x,y)$ is just the Faddeev-Popov determinant, and
therefore $\Delta \ne 0.$ Hence, one has from Eq.~(\ref{cderiv})
$F_{\alpha}\xi^{\alpha\beta} = 0,$ and, in view of non-degeneracy
of $\xi^{\alpha\beta},$ $F_{\alpha} = 0.$ The classical metric is
thus independent of the choice of the matrix $\xi^{\alpha\beta},$
and in particular, of the replacements $F_{\alpha} \to
A_{\alpha}^{\beta}F_{\beta}.$ One can put this in another way by
saying that the weighting matrix has no geometrical meaning in
classical theory.

This differs, however, in quantum domain. The classical equations
(\ref{classeq}) are replaced in quantum theory by the effective
equations
$$\frac{\delta \Gamma}{\delta g^{\rm eff}_{\mu\nu}} =
- T_{\rm eff}^{\mu\nu}\,,$$ where $\Gamma,$ $g^{\rm
eff}_{\mu\nu},$ and $T_{\rm eff}^{\mu\nu}$ are the effective
action, metric, and energy-momentum tensor of matter,
respectively. In general, the fields $g^{\rm eff}_{\mu\nu}$ do not
satisfy the gauge conditions $F_{\alpha} = 0,$ and moreover,
depend on the choice of the weighting matrix $\xi^{\alpha\beta};$
$\xi^{\alpha\beta}$-independence is inherited only by the tree
contribution.

Dependence on the choice of the weighting matrix generally
represents an excess of the gauge arbitrariness over the
arbitrariness in the choice of reference frame; it is therefore a
potential source of ambiguity in the values of observables. This
dependence causes no gauge ambiguity of observables only if it
reduces to the symmetry transformations. In other words, under
(infinitesimal) variations of the matrix $\xi^{\alpha\beta},$ the
fields $g^{\rm eff}_{\mu\nu}$ must transform as in
Eq.~(\ref{gaugesym})
\begin{eqnarray}\label{gaugesymeff}
\delta g^{\rm eff}_{\mu\nu} = \Xi^{\alpha}g^{\rm
eff}_{\mu\nu,\alpha} + g^{\rm eff}_{\mu\alpha}\Xi^{\alpha}_{,\nu}
+ g^{\rm eff}_{\nu\alpha}\Xi^{\alpha}_{,\mu}\,,
\end{eqnarray}
\noindent with some functions $\Xi^{\alpha}.$ It will be shown in
the following Section that this is the case indeed.

\paragraph{The one-loop level.}\label{oneloop}

\begin{figure}
\scalebox{0.8}{\includegraphics{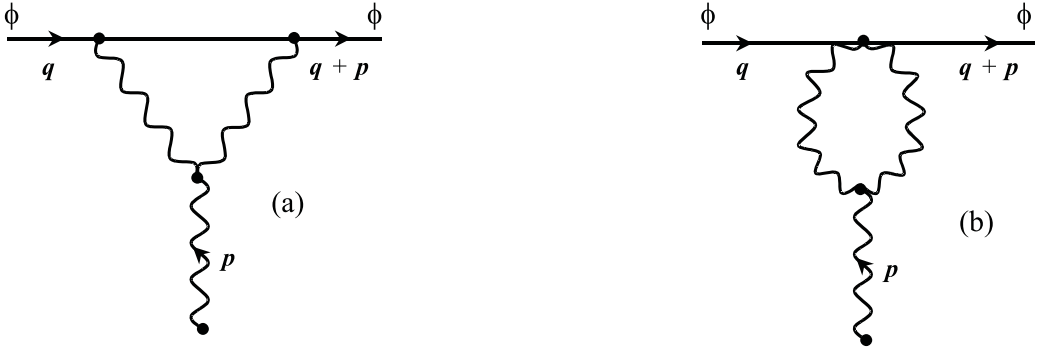}} \caption{Some of the
one-loop diagrams contributing to the effective gravitational
field. (a) The only diagram giving rise to the root singularity
with respect to the momentum transfer $p.$ (b) An example of a
diagram free of the root singularity. Wavy lines represent
gravitons, solid lines massive particles. $q$ is the particle
4-momentum.} \label{fig2}
\end{figure}

Let us now turn to the examination of the gauge dependence of
$\hbar^0$ loop contribution to the effective gravitational field.
This contribution comes from diagrams in which virtual propagation
of matter fields is near their mass shells, and is represented by
terms containing the root singularity with respect to the momentum
transfer between gravitational and matter fields. In the first
post-Newtonian approximation, the only diagram we need to consider
is the one-loop diagram depicted in Fig.~\ref{fig2}(a). As a
simple analysis shows, other one-loop diagrams do not contain the
root singularities, while many-loop diagrams are of higher orders
in the Newton constant. For instance, the loop in the diagram of
Fig.~\ref{fig2}(b) does not contain massive particle propagators,
and therefore, expands in integer powers (or logarithms) of the
ratio $\alpha = -p^2/m^2.$

\begin{figure}
\scalebox{0.65}{\includegraphics{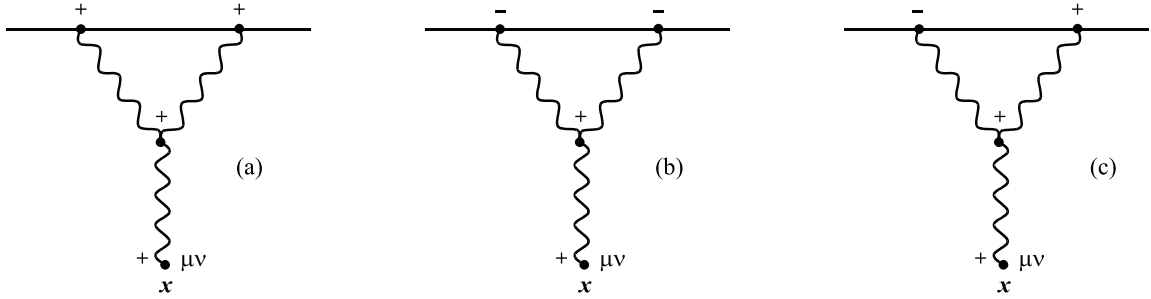}} \caption{Some of
diagrams arising upon writing out the matrix diagram in
Fig.~\ref{fig2}(a) according to the Schwinger-Keldysh rules. (a)
The only diagram giving rise to a nonzero contribution in the
long-range limit. (b) Diagram vanishing in the long-range limit
because of the condition $p^0 = 0.$ (c) An example of a diagram
which is zero identically because of the vanishing of one of its
vertices (the left $\phi^2 h$ vertex).} \label{fig3}
\end{figure}

Let us show, first of all, that the Schwinger-Keldysh rules
applied to the diagram in Fig.~\ref{fig2}(a) reduce to the
ordinary Feynman rules of the standard S-matrix theory. According
to the former, there are eight diagrams corresponding to the eight
different ways of placing the three internal vertices of this
diagram into the plus and minus branches of the time path (see
Fig.~\ref{fig3}). But all diagrams in Fig.~\ref{fig3} except the
first two diagrams [2(a) and 2(b)] are zeros identically because
of the energy-momentum conservation in the vertices, and the mass
shell condition for external matter lines (massive particle cannot
emit a real massless graviton). As to the diagram 2(b), its
$\hbar^0$ order contribution is also zero. Indeed, the time
component of the momentum transfer
$$p^0 = \sqrt{m^2 + (\bm{p} + \bm{q})^2} - \sqrt{m^2 + \bm{q}^2} =
O\left(\frac{\bm{p q}}{m}\right)\,.$$ To the leading order,
therefore, $p^0$ is to be set zero, and the product of the two
internal graviton propagators vanishes because the arguments of
the corresponding $\theta$-functions are opposite:
$$D_{-+}(k)D_{+-}(k+p) \sim \theta(k^0)\theta(-k^0) = 0\,.$$
The remaining diagram 2(a) contains only Feynman (causal)
propagators, and therefore so do all other diagrams arising in the
derivations of this section. Accordingly, we will omit the
subscripts ${\rm SK},$ $\pm,$ and denote the $D_{++}$ functions
simply by $D.$

\begin{figure}
\includegraphics{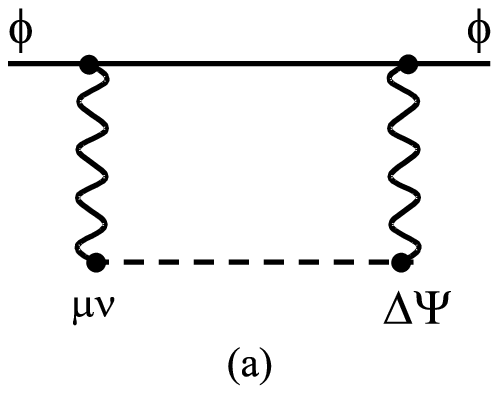}
\includegraphics{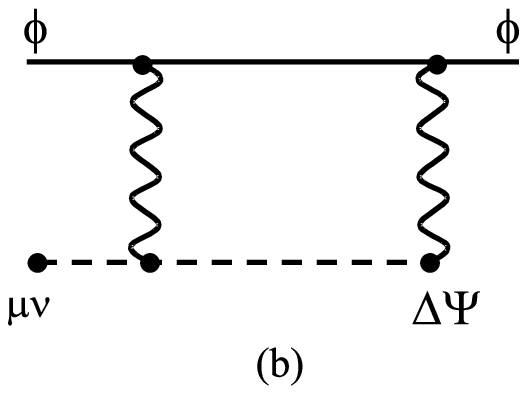}
\includegraphics{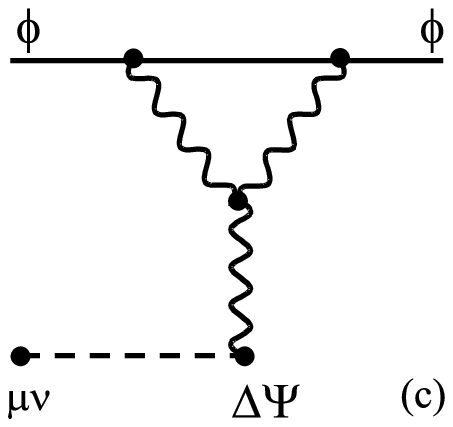}
\includegraphics{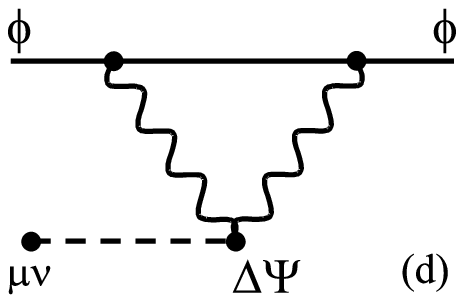}
\caption{The one-loop diagrams giving rise to the root singularity
in the right hand side of Eq.~(\ref{hvar1}). Dashed lines
represent the Faddeev-Popov ghosts.} \label{fig4}
\end{figure}

It follows from Eq.~(\ref{wvar}) that under a variation
$\Delta\Psi$ of the gauge fermion, variation of the effective
gravitational field
$$h^{\rm eff}_{\mu\nu} = \frac{\delta W}{\delta t^{\mu\nu}}$$
has the form
\begin{eqnarray}&&\label{hvar}
\delta h^{\rm eff}_{\mu\nu} = \left( \frac{\delta}{\delta
k^{\mu\nu}}\langle \Delta\Psi\rangle + j^i\frac{\delta^2}{\delta
t^{\mu\nu}\delta q^i} \langle \Delta\Psi\rangle
\right)_{\genfrac{}{}{0pt}{}{J\setminus j = 0}{K = 0}}\,,
\end{eqnarray}
\noindent where $J\setminus j$ means that the source $j$ is
excluded from $J.$ We are interested presently in variations of
the weighting matrix $\xi^{\mu\nu},$ therefore,
$$\Delta\Psi(\Phi,K)
= -
\frac{\bar{c}^{\alpha}}{2}\pi^{\beta}\Delta\zeta_{\beta\alpha}\,,$$
or, integrating $\pi^{\alpha}$ out,
\begin{eqnarray}&&\label{psivar}
\Delta\Psi(\Phi,K) = \frac{\bar{c}^{\alpha}}{2}\zeta_{\alpha\beta}
\Delta\xi^{\beta\gamma}F_{\gamma}\,.
\end{eqnarray}
\noindent According to general rules, in order to find the
contribution of a diagram with $n$ external $\phi$-lines, one has
to take the $n$th derivative of the right hand side of
Eq.~(\ref{hvar}) with respect to $j^i,$ multiply the result by the
product of $n$ factors $e_i(q^2 - m^2),$ where $q, e_i$ are the
4-momentum and polarization of the external $\phi_i$-field quanta,
and set $q^2 = m^2$ afterwards. The second term on the right of
Eq.~(\ref{hvar}) is proportional to the source $j^i$ contracted
with the vertex $G_i^{\alpha}c_{\alpha}.$ This term represents
contribution of the graviton propagators ending on the external
matter lines. Multiplied by $(q^2 - m^2),$ it gives rise to a
non-zero value as ${q^2\to m^2}$ only if the corresponding diagram
is one-particle-reducible with respect to the $\phi$-line, in
which case it describes the variation of $h^{\rm eff}_{\mu\nu}$
under the gauge variation of external matter lines. It is
well-known, however, that $\phi$-operators must be
renormalized\footnote{One might think that the gauge dependence of
the renormalization constants could spoil the above derivation of
Eq.~(\ref{wvar}). In fact, this equation holds true for
renormalized as well as unrenormalized quantities \cite{tyutin}.}
so as to cancel all the radiative corrections to the external
lines\footnote{The above discussion is nothing but the well-known
reasoning underlying the proof of gauge-independence of the
$S$-matrix \cite{tyutin2}.}. Therefore, this term can be omitted,
and Eq.~(\ref{hvar}) rewritten finally as
\begin{eqnarray}&&\label{hvar1}
\delta h^{\rm eff}_{\mu\nu} = \frac{1}{2}\zeta_{\alpha\beta}
\Delta\xi^{\beta\gamma} \left.\frac{\delta\langle\bar{c}^{\alpha}
F_{\gamma}\rangle} {\delta
k^{\mu\nu}}\right|_{\genfrac{}{}{0pt}{}{J\setminus j = 0}{K =
0}}\,.
\end{eqnarray}
\noindent The one-loop diagrams representing the right hand side
of Eq.~(\ref{hvar1}), which give rise to the root singularity, are
shown in Fig.~\ref{fig4}. Let us consider the diagram of
Fig.~\ref{fig4}(a) first. It turns out that this diagram is
actually free of the root singularity despite the presence of the
internal $\phi$-line. The rightmost vertex in this diagram is
generated by $\bar{c}^{\alpha}F^{(1)}_{\gamma},$ where
$F^{(1)}_{\gamma}$ denotes the linear part of $F_{\gamma}.$ The
graviton propagator connecting this vertex to the $\phi$-line can
be expressed through the ghost propagator with the help of the
equation
\begin{eqnarray}\label{slavtree}
\xi^{\alpha\beta}F_{\beta}^{(1),\sigma\lambda}
D_{\sigma\lambda\mu\nu} =
G^{(0)\beta}_{\mu\nu}\tilde{D}_{\beta}^{\alpha}\,, \qquad
G^{(0)\beta}_{\mu\nu}\equiv
\left.G^{\beta}_{\mu\nu}\right|_{h=0}\,,
\end{eqnarray}
\noindent where $\tilde{D}_{\beta}^{\alpha}$ is the ghost
propagator defined by
\begin{eqnarray}&&\label{ghostp}
F_{\alpha}^{(1),\mu\nu}G^{(0)\beta}_{\mu\nu}\tilde{D}^{\gamma}_{\beta}
= - \delta_{\alpha}^{\gamma}\,,
\end{eqnarray}
\noindent respectively. Equation (\ref{slavtree}) is the Slavnov
identity (\ref{slav}) at the tree level, differentiated twice with
respect to $t^{\mu\nu},$ $\beta_{\alpha}.$ Using this identity in
the diagram Fig.~\ref{fig4}(a) we see that the ghost propagator is
attached to the matter line through the generator
$D^{(0)\alpha}_{\mu\nu}.$ On the other hand, the action $S_{\phi}$
is invariant under the gauge transformations (\ref{gaugesymm}) --
(\ref{gaugesymmsigma}),
\begin{eqnarray}&&\label{gaugesym1}
\frac{\delta S_{\phi}}{\delta\phi_i}G_i^{\alpha} + \frac{\delta
S_{\phi}}{\delta h_{\mu\nu}}G^{\alpha}_{\mu\nu} = 0.
\end{eqnarray}
\noindent Differentiating Eq.~(\ref{gaugesym1}) twice with respect
to $\phi,$ and setting $h_{\mu\nu} = 0$ yields
\begin{eqnarray}&&\label{ident1}
\frac{\delta S^{(2)}_{\phi}}{\delta\phi_i
\delta\phi_k}\frac{\delta G_i^{\gamma}}{\delta\phi_l} +
\frac{\delta S^{(2)}_{\phi}}{\delta\phi_i
\delta\phi_l}\frac{\delta G_i^{\gamma}}{\delta\phi_k} +
\left.\frac{\delta^3 S_{\phi}}{\delta
h_{\sigma\lambda}\delta\phi_l
\delta\phi_k}\right|_{h=0}G^{(0)\gamma}_{\sigma\lambda} = 0\,.
\end{eqnarray}
\noindent Taking into account also the mass shell condition
$$\frac{\delta^2 S^{(2)}_{\phi}}{\delta\phi_i \delta\phi_k}\bar{\phi}_k = 0,$$
we see that under contraction of the vertex factor with the
external and internal matter lines, the $\phi$-particle
propagator, $D_{ik},$ satisfying
$$\frac{\delta^2 S^{(2)}_{\phi}}{\delta\phi_i\delta\phi_k}
D_{kl} = - \delta_{l}^{i}\,,$$ is cancelled
\begin{eqnarray}\label{cancel}
D_{kl}\left.\frac{\delta^2 S^{(2)}_{\phi}}{\delta\phi_k\delta
h_{\mu\nu}}\right|_{h=0}G^{(0)\alpha}_{\mu\nu} =  G_l^{\alpha}\,.
\end{eqnarray}
\noindent We conclude that the $\hbar^0$ contribution of the
diagram Fig.~\ref{fig4}(a) is zero. As to the rest of diagrams,
they are all proportional to the generator
$D^{(0)\alpha}_{\mu\nu}.$ Thus, the right hand side of
Eq.~(\ref{hvar1}) can be written
\begin{eqnarray}
\delta h^{\rm eff}_{\mu\nu} = G^{(0)\alpha}_{\mu\nu}\Xi_{\alpha} +
O(\hbar)\,, \qquad \Xi_{\alpha} =
\frac{1}{2}\tilde{D}_{\alpha}^{\beta}
\zeta_{\beta\gamma}\Delta\xi^{\gamma\delta}\langle
F_{\delta}\rangle \,. \nonumber
\end{eqnarray}
\noindent Since $\Xi_{\alpha}$ are of the order $G^2,$ one can
also write, within the accuracy of the first post-Newtonian
approximation,
\begin{eqnarray}\label{mainaux}
\delta h^{\rm eff}_{\mu\nu} = G^{\alpha}_{\mu\nu}\Xi_{\alpha}\,,
\end{eqnarray}
\noindent where $G^{\alpha}_{\mu\nu}$ are defined by
Eq.~(\ref{gaugesymmh}) with $h_{\mu\nu} \to h^{\rm
eff}_{\mu\nu}\,.$

We thus see that under variations of the weighting matrix, the
effective metric does transform according to
Eq.~(\ref{gaugesymeff}). To determine the effect of these
variations on the values of observables, one has to find also the
induced transformation of the reference frame, {\it i.e.}, of the
functions $\sigma_a.$ Obviously, the gauge variation of
$\sigma_a$'s is represented by the same set of diagrams pictured
in Fig.~\ref{fig4}(b,c,d),\footnote{The diagram of
Fig.~\ref{fig4}(a) does not contribute in this case.} with the
only difference that the leftmost vertex ($\mu\nu$) in these
diagrams is now generated by $\sigma_{a,\alpha}c^{\alpha}$ instead
of $D^{\alpha}_{\mu\nu}c_{\alpha}.$ Thus, under variations of the
weighting matrix, the functions $\sigma_a(x)$ transform according
to
\begin{eqnarray}\label{mainaux1}
\delta \sigma_a = \sigma_{a,\alpha}\Xi^{\alpha}\,,
\end{eqnarray}
\noindent where $\Xi^{\alpha}$'s are the same as in
Eq.~(\ref{mainaux}).

Equations (\ref{mainaux}) and (\ref{mainaux1}) are of the same
form as Eqs.~(\ref{vartreeg}) and (\ref{vartreesigma}),
respectively, which implies that the value of any observable $O$
is invariant under variations of the weighting matrix,
\begin{eqnarray}\label{example1}
\delta O[h^{\rm eff}(x(\sigma^0))] = \frac{\delta O}{\delta h^{\rm
eff}_{\mu\nu}}\delta h^{\rm eff}_{\mu\nu} + \frac{\partial
O}{\partial x^{\mu}}\delta x^{\mu} = \frac{\partial O}{\partial
x^{\alpha}}\Xi^{\alpha} - \frac{\partial O}{\partial
x^{\mu}}\Xi^{\mu} = 0\,.
\end{eqnarray}
\noindent In particular, $$\delta R[g^{\rm eff}(x(\sigma^0))] =
0.$$ Furthermore, any tensor quantity $O_{\alpha\beta...}$ (or
$O^{\mu\nu...}$), calculated at a fixed reference point
$\sigma^0,$ transforms covariantly (contravariantly), as
prescribed by the position of the tensor indices of the
corresponding operator. This is in accord with the principle of
general covariance.

\subsubsection{Dependence of effective metric on the form of
$F_{\alpha}$}\label{general}

Having established the general law of the effective metric
transformation under variations of the weighting matrix, let us
turn to investigation of the variations of the functions
$F_{\alpha}$ themselves.

According to the general equation (\ref{hvar}), a variation
$\Delta F_{\alpha}$ induces the following variation in the
effective metric
\begin{eqnarray}&&\label{hvarf}
\delta h^{\rm eff}_{\mu\nu} =
\left.\frac{\delta\langle\bar{c}^{\alpha}\Delta F_{\alpha}\rangle}
{\delta k^{\mu\nu}}\right|_{\genfrac{}{}{0pt}{}{J\setminus j =
0}{K = 0}}\,.
\end{eqnarray}
\noindent The general structure of diagrams representing the
one-loop contribution to the right hand side of this equation is
the same as before and given by Fig.~\ref{fig4}. Contribution of
diagrams (b), (c), and (d) is again a spacetime diffeomorphism. In
the present case, however, diagram (a) gives rise to a non-zero
contribution already in the order $\hbar^0.$ Namely, it is not
difficult to show that the combination $\Delta
F^{(1),\mu\nu}_{\alpha} D_{\mu\nu\sigma\lambda}$ cannot be brought
to the form proportional to the generator $G^{(0)}.$ Note, first
of all, that the variation of $\Delta F^{(1),\mu\nu}_{\alpha}
D_{\mu\nu\sigma\lambda}$ with respect to $\xi^{\alpha\beta}$ {\it
is} proportional to $G^{(0)};$ in the highly condensed DeWitt's
notation,
$$\delta (\Delta F^{(1)}_1 D) =
\Delta F^{(1)}_1 D (F^{(1)}_{1}\delta\xi F^{(1)}_1) D = \Delta
F^{(1)}_1 D F^{(1)}_1\delta\xi \zeta \tilde{D} G^{(0)},$$ where
the Slavnov identity (\ref{slavtree}) has been used. Hence,
without changing the $\hbar^0$ part of diagram (a),
$\zeta_{\alpha\beta}$ can be set zero, in which case
Eq.~(\ref{slavtree}) gives $F^{(1)}_1 D = 0.$ Suppose that $\Delta
F^{(1),\mu\nu}_{\alpha} D_{\mu\nu\sigma\lambda} =
X_{\alpha\beta}G^{(0)\beta}_{\sigma\lambda},$ or shorter, $\Delta
F^{(1)}_1 D = X G^{(0)},$ with some $X.$ Then one has $0 = \Delta
F^{(1)}_1 D F^{(1)}_1 = X F^{(1)}_1 G^{(0)}\equiv X M(h=0).$ Since
the Faddeev-Popov determinant $\det M \ne 0,$ it follows that $X =
0.$ Thus, the argument used in the preceding section does not
work, and the question is whether contribution of the diagram (a)
can be actually represented in the form
$G_{\mu\nu}^{\alpha}\Xi_{\alpha}.$

The answer to this question is negative, as an explicit
calculation shows. This will be demonstrated below in the simplest
case of a single scalar field described by the action
\begin{eqnarray}&&\label{actionm}
S_{\phi} =  \frac{1}{2}{\displaystyle\int} d^4 x
\sqrt{-g}\left\{g^{\mu\nu}\partial_{\mu}\phi \partial_{\nu}\phi -
m^2\phi^2\right\},
\end{eqnarray}
\noindent and linear gauge conditions
\begin{eqnarray}&&\label{gauge}
F_{\gamma} = \eta^{\mu\nu}\partial_{\mu}h_{\nu\gamma} -
\left(\frac{\varrho - 1}{\varrho - 2}\right)\partial_{\gamma}h\,,
\qquad h \equiv \eta^{\mu\nu}h_{\mu\nu}\,, \qquad
\zeta_{\alpha\beta} = 0\,,
\end{eqnarray}
\noindent where $\varrho$ is an arbitrary parameter. According to
Eq.~(\ref{hvarf}), the $\varrho$-derivative of the effective
metric is given by
\begin{eqnarray}&&\label{hvarfa}
\frac{\partial h^{\rm eff}_{\mu\nu}}{\partial\varrho} =
\left.\frac{\delta\langle\bar{c}^{\alpha}
\partial F_{\alpha}/\partial\varrho\rangle}
{\delta k^{\mu\nu}}\right|_{\genfrac{}{}{0pt}{}{J\setminus j =
0}{K = 0}}\,.
\end{eqnarray}
\noindent

\begin{figure}
\scalebox{0.8}{\includegraphics{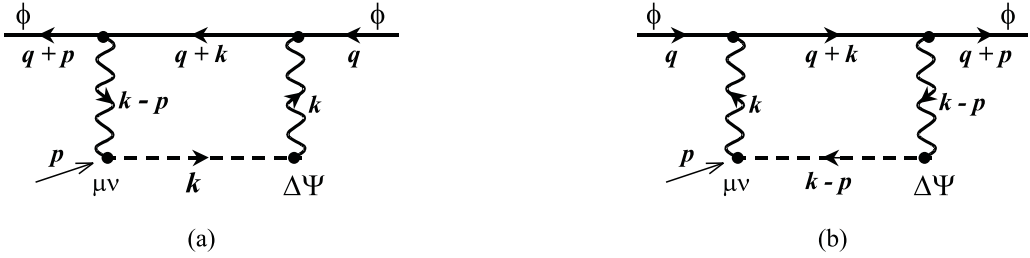}} \caption{Diagrams
responsible for the nontrivial contribution to the right hand side
of Eq.~(\ref{hvarfa}).} \label{fig5}
\end{figure}

There are two diagrams with the structure of Fig.~\ref{fig4}(a),
in which the scalar particle propagates in opposite directions.
They are represented in Fig.~\ref{fig5}. In fact, it is sufficient
to evaluate either of them. Indeed, these diagrams have the
following tensor structure $$a_1 q_{\mu}q_{\nu} + a_2
(p_{\mu}q_{\nu} + p_{\nu}q_{\mu}) + a_3 p_{\mu}p_{\nu} +
a_4\eta_{\mu\nu},$$ where $a_i, i = 1,...,4,$ are some functions
of $p^2.$ When transformed to the coordinate space, the second and
third terms become spacetime gradients, hence, they can be written
in the form $G_{\mu\nu}^{(0)\alpha}\Xi_{\alpha}.$ As was discussed
in the preceding sections, the terms of this type respect general
covariance, therefore, we can restrict ourselves to the
calculation of $a_1$ and $a_4$ only. On the other hand, diagrams
of Fig.~\ref{fig5} go over one into another under the substitution
$q \to - q - p$ which leaves $a_1, a_4$ unchanged. Thus, we have
\begin{eqnarray}\label{int}
I_{\mu\nu}(x) &=& \iint \frac{d^3 \bm{q}}{(2\pi)^3} \frac{d^3
\bm{p}}{(2\pi)^3}\frac{a(\bm{q})a^*(\bm{q} +
\bm{p})}{\sqrt{2\varepsilon_{\bm q}2\varepsilon_{{\bm q} +
\bm{p}}}} e^{ipx}\tilde{I}_{\mu\nu}(p,q)\,, \quad p_0 =
\varepsilon_{\bm{q} + \bm{p}} - \varepsilon_{\bm{q}}\,,
\\
\tilde{I}_{\mu\nu}(p,q) &=& \tilde{I}^{3(a)}_{\mu\nu}(p,q) +
\tilde{I}^{3(b)}_{\mu\nu}(p,q)\,, \nonumber\\
\tilde{I}^{3(a)}_{\mu\nu}(p,q) &=& -
i\mu^{\epsilon}{\displaystyle\int} \frac{d^{4-\epsilon}
k}{(2\pi)^4} \left\{ \frac{1}{2} W^{\alpha\beta\gamma\delta}
(q_{\gamma} + p_{\gamma}) (q_{\delta} + k_{\delta}) - m^2
\frac{\eta^{\alpha\beta}}{2} \right\} \nonumber\\&& \times D(q +
k)\left\{ \frac{1}{2} W^{\rho\tau\sigma\lambda} q_{\sigma}
(q_{\lambda} + k_{\lambda}) - m^2
\frac{\eta^{\rho\tau}}{2}\right\} D_{\rho\tau\pi\omega}(k)
\frac{\partial F^{\xi,\pi\omega}}{\partial\varrho} \nonumber\\&&
\times\tilde{D}_{\xi}^{\zeta}(k) \left\{-(k_{\zeta} - p_{\zeta})
\delta_{\mu\nu}^{\chi\theta} + \delta_{\zeta\mu}^{\chi\theta}
k_{\nu} + \delta_{\zeta\nu}^{\chi\theta} k_{\mu} \right\}
D_{\chi\theta\alpha\beta}(k - p)\,, \nonumber\\
\tilde{I}^{3(b)}_{\mu\nu}(p,q) &=&
\tilde{I}^{3(a)}_{\mu\nu}(p,-q-p)\,,\nonumber
\end{eqnarray}
\noindent where $a(\bm{q})$ is the momentum space amplitude for
the scalar particle, normalized by
$$\int\frac{d^3 \bm{q}}{(2\pi)^3}|a(\bm{q})|^2 = 1\,,$$
$\mu$ is an arbitrary mass scale, $\varepsilon_{\bm{q}} =
\sqrt{\bm{q}^2 + m^2}\ ,$ and $\epsilon = 4 - d,$ $d$ being the
dimensionality of spacetime. Explicit expressions for the
propagators
\begin{eqnarray}
D_{\mu\nu\sigma\lambda}(k) &=& \frac{W_{\mu\nu\sigma\lambda}}{k^2}
- \varrho (\eta_{\mu\nu}k_{\sigma}k_{\lambda} +
\eta_{\sigma\lambda}k_{\mu}k_{\nu})\frac{1}{k^4}
\nonumber\\
&+& (\eta_{\mu\sigma} k_{\nu} k_{\lambda} + \eta_{\mu\lambda}
k_{\nu} k_{\sigma} + \eta_{\nu\sigma} k_{\mu} k_{\lambda} +
\eta_{\nu\lambda} k_{\mu} k_{\sigma}) \frac{1}{k^4}
\nonumber\\
&+& (3\varrho^2 - 4\varrho)k_{\mu}k_{\nu}
k_{\sigma}k_{\lambda}\frac{1}{k^6}\,,
\label{hprop}\\
\tilde{D}^{\alpha}_{\beta}(k) &=&
\frac{\delta^{\alpha}_{\beta}}{k^2} -
\frac{\varrho}{2}\frac{k^{\alpha}k_{\beta}}{k^4}\,,
\nonumber\\
D(k) &=&  \frac{1}{m^2 - k^2}\,, \nonumber
\end{eqnarray}
\noindent Calculation of (\ref{int}) can be further simplified
using the relation
$$\frac{\partial F_1}{\partial\varrho} D
= - F_1 \frac{\partial D}{\partial\varrho}\,,$$ which follows from
$F_1 D = 0,$ and noting that all gradient terms in the graviton
propagators, contracted with the $\phi^2 h$ vertices, can be
omitted (see Sec.~\ref{feynman}), i.e., only the first line in
Eq.~(\ref{hprop}) actually contributes.

Upon extraction of the leading contribution, Eq.~(\ref{int})
considerably simplifies. Note, first of all, that to the leading
order in the long-range expansion, one has $\varepsilon_{\bm{q} +
\bm{p}}\approx \varepsilon_{\bm{q}} \approx m,$ and hence, $p_0
\approx 0.$ Next, take into account that the function $a(\bm{q})$
is generally of the form
$$a(\bm{q}) = b(\bm{q}) e^{-i\bm{q}\bm{x}_0},$$ where $\bm{x}_0$
is the mean particle position, and $b(\bm{q})$ is such that
\begin{eqnarray}\label{packet}
\int d^3\bm{q}~b(\bm{q})e^{i\bm{qx}} = 0
\end{eqnarray}
\noindent for $\bm{x}$ outside of some finite region $W$ around
$\bm{x} = 0.$ In the long-range limit, $b(\bm{q} + \bm{p})$ may be
substituted by $b(\bm{q}):$ This implies that we disregard spatial
spreading of the wave packet, neglecting the multipole moments of
the particle mass distribution. Hence, Eq.~(\ref{int}) can be
written as
\begin{eqnarray}\label{intnew} I_{\mu\nu}(x)
&=& \frac{1}{2m}\iint \frac{d^3 \bm{q}}{(2\pi)^3} \frac{d^3
\bm{p}}{(2\pi)^3}|b(\bm{q})|^2 e^{-i\bm{p}(\bm{x} - \bm{x}_0
)}\tilde{I}_{\mu\nu}(p,q)\,, \quad p^0 = 0\,.
\end{eqnarray}
\noindent

Let the equality of two functions up to a spacetime diffeomorphism
be denoted by ``$\sim$''. Then, performing tensor multiplications
in Eq.~(\ref{int}), and omitting terms proportional to $p_{\mu},$
one obtains
\begin{eqnarray}&&
\tilde{I}_{\mu\nu}^{3(a)}(p,q) \sim - i\mu^{\epsilon}
{\displaystyle\int} \frac{d^{4-\epsilon} k}{(2\pi)^4}
\frac{1}{k^4}\frac{1}{(k + p)^2}\frac{1}{m^2 - (k + q)^2}
\nonumber\\&& \left\{ \eta_{\mu\nu}\left[ \frac{\varrho }{2} (k^4
+ 2 P Q) (P - Q) (Q - m^2) \right. \right. \nonumber\\&& \left.
\left. + \frac{\varrho }{2} k^2 P (P + Q) (Q - m^2)) - \varrho k^2
Q^2 (Q - m^2) \right. \right. \nonumber\\&& \left. \left. +
\left(\frac{\varrho}{4} p^2 (k^2 + 2 Q) + (k + p)^2 m^2
\right)(k^2 + P) (Q - m^2) \right] \right. \nonumber\\&& \left. +
k_{\mu} k_{\nu} \left[ \varrho (P^2 + k^2 m^2) (Q - m^2) + 4
\varrho P Q m^2 - 3 \varrho P Q^2 \right. \right. \nonumber\\&&
\left. \left. + \frac{\varrho}{2} p^2 (P - 2 Q) (Q - m^2) + 2
\varrho Q^2 (Q - m^2) - \varrho P m^4 \right. \right.
\nonumber\\&& \left. \left. + 2 (k + p)^2 (Q (Q - 2 m^2) - P (Q -
m^2) + m^4) \right] \right. \nonumber\\&& \left. + 2 (k_{\mu}
q_{\nu} + k_{\nu} q_{\mu})(k + p)^2 (Q - P) (Q - m^2) \right.
\nonumber\\&& \left. - 2 q_{\mu} q_{\nu} (k + p)^2 (k^2 + P) (Q -
m^2) \right\}\,, \qquad Q\equiv (kq), \quad P\equiv (kp).
\end{eqnarray}
\noindent Introducing the Schwinger parameterization of
denominators
\begin{eqnarray}&&
\frac{1}{k^2} = - \int_{0}^{\infty} dy\exp\{y k^2\}\,,
\qquad\frac{1}{(k + p)^2} = - \int_{0}^{\infty} dx\exp\{x (k +
p)^2\}\,, \nonumber\\&& \frac{1}{k^2 + 2 (kq)} = -
\int_{0}^{\infty} dz \exp\{z [k^2 + 2 (kq)]\}\,, \nonumber
\end{eqnarray}
one evaluates the loop integrals using
\begin{eqnarray}&&
\int~d^{d}k \exp\{ k^2 (x + y + z) + 2 k^{\mu} (x p_{\mu} + z
q_{\mu}) + p^2 x \} \nonumber\\&& = i \left(\frac{\pi}{x + y +
z}\right)^{d/2} \exp\left\{\frac{p^2 x y - m^2 z^2}{x + y +
z}\right\}, \nonumber
\end{eqnarray}
\begin{eqnarray}&&
\int~d^{d}k ~k_{\alpha} \exp\{ k^2 (x + y + z) + 2 k^{\mu} (x
p_{\mu} + z q_{\mu}) + p^2 x \} = \nonumber\\&& = i
\left(\frac{\pi}{x + y + z}\right)^{d/2} \exp\left\{\frac{p^2 x y
- m^2 z^2}{x + y + z}\right\} \left[- \frac{x p_{\alpha} + z
q_{\alpha}}{x + y + z}\right], \nonumber
\end{eqnarray}
\noindent etc., up to six $k$-factors in the integrand. This
calculation can be automated to a considerable extent with the
help of the tensor package \cite{reduce} for the REDUCE system.
Changing the integration variables $(x,y,z)$ to $(t,u,v)$ via
$$x = \frac{t (1 + t + u) v}{m^2 (1 + \alpha t u)}\,,
\qquad y = \frac{u (1 + t + u) v}{m^2 (1 + \alpha t u)}\,, \qquad
z = \frac{ (1 + t + u) v}{m^2 (1 + \alpha t u)}\,, \qquad \alpha
\equiv - \frac{p^2}{m^2}\,,$$ integrating $v$ out, subtracting the
ultraviolet divergence\footnote{ A technicality must be mentioned
here. By itself, the diagram of Fig.~\ref{fig3}(a) is free of
infrared divergences. As a result of the BRST-operating with this
diagram, however, some fictitious infrared divergences are brought
into individual diagrams representing the right hand side of
Eq.~(\ref{hvar}). This is because the vertex
$D^{\alpha}_{\mu\nu}C_{\alpha}$ contains the term
$C^{\alpha}\partial_{\alpha}h_{\mu\nu}$ in which the spacetime
derivatives act on the gravitational, rather than the ghost field.
These divergences occur as $u,t \to \infty.$ They are proportional
to integer powers of $p^2,$ and therefore do not interfere with
the part containing the root singularity. Since these divergences
must eventually cancel in the total sum in Eq.~(\ref{hvar}), they
will be simply omitted in what follows.}
$$\tilde{I}_{\mu\nu}^{3(a){\rm div}}(p,q) = - \frac{1}{16\pi^2 \epsilon}
\left(\frac{\mu}{m}\right)^{\epsilon} \left[\frac{1}{3} q_{\mu}
q_{\nu} + \eta_{\mu\nu}(p^2 - 2 m^2) \frac{3\varrho - 2}{24}
\right],$$ setting $\epsilon = 0$, omitting gradient terms, and
retaining only the $\hbar^0$-contribution, we obtain
\begin{eqnarray}\label{int2}
\tilde{I}^{3(a)\rm ren}_{\mu\nu}(p,q) &\equiv&
(\tilde{I}_{\mu\nu}^{3(a)} - \tilde{I}_{\mu\nu}^{3(a){\rm
div}})_{{\epsilon} \to 0}
\nonumber\\
&\sim&  \frac{m^2}{16\pi^2} \int_{0}^{\infty}\int_{0}^{\infty} du
dt \left\{ \frac{\eta_{\mu\nu}\varrho}{H^3 \alpha}\left(
\frac{1}{D^2} - \frac{1}{2 D}\right) \right. \nonumber\\&+& \left.
\frac{q_{\mu} q_{\nu}}{H^3 m^2}\left[ \frac{2 H^2}{D^2}\left(1 -
\frac{1}{D}\right) + \frac{1}{\alpha}\left(\frac{4\varrho}{D^3} -
\frac{11\varrho + 4}{D^2} + \frac{8\varrho + 4}{D}\right) \right]
\right\}\,, \nonumber\\&& D \equiv 1 + \alpha u t\,, \qquad H
\equiv 1 + u + t\,.
\end{eqnarray}
\noindent The root singularity in the right hand side of
Eq.~(\ref{int2}) can be extracted using Eqs.~(\ref{roots}) derived
in the Appendix. Denoting
$$\int_{0}^{\infty}\int_{0}^{\infty}
du dt~H^{- n} D^{- m} \equiv J_{nm} \,,$$ one has
\begin{eqnarray}
J^{{\rm root}}_{12} &=& \frac{\pi^2}{4\sqrt{\alpha}}\,, ~~J^{{\rm
root}}_{13} = \frac{3\pi^2}{16\sqrt{\alpha}}\,,
\nonumber\\
J^{{\rm root}}_{31} &=& - \frac{\pi^2}{16}\sqrt{\alpha}\,,
~~J^{{\rm root}}_{32} = - \frac{3\pi^2}{32}\sqrt{\alpha}\,,
~~J^{{\rm root}}_{33} = - \frac{15\pi^2}{128}\sqrt{\alpha}\,.
\nonumber
\end{eqnarray}
\noindent Substituting these into Eq.~(\ref{int2}) gives
\begin{eqnarray}&&\label{int3new}
\tilde{I}^{\rm ren}_{\mu\nu}(p,q) \sim \frac{1}{128\sqrt{\alpha}}
[q_{\mu} q_{\nu}(\varrho + 1) - \eta_{\mu\nu} m^2 \varrho].
\end{eqnarray}\noindent
Finally, restoring ordinary units, substituting
Eq.~(\ref{int3new}) into Eq.~(\ref{intnew}), and using the formula
\begin{eqnarray}\label{rootint}
\int \frac{d^3\bm{p}}{(2\pi)^3}\frac{e^{i \bm{p x}}}{|\bm{p}|} =
\frac{1}{2\pi^2 r^2}\,, \quad r^2 \equiv \delta_{i k }x^i x^k \,,
\end{eqnarray}
\noindent we obtain the following expression for the
$\varrho$-derivative of the $G^2\hbar^0$-order contribution to the
effective metric \cite{kazakov6}
\begin{eqnarray}&&\label{mainnew}
\frac{\partial h^{\rm eff}_{\mu\nu}}{\partial\varrho} =
\frac{\partial h^{\rm tree}_{\mu\nu}}{\partial\varrho} +
\frac{\partial h^{\rm loop}_{\mu\nu}}{\partial\varrho} \sim
\frac{\partial h^{\rm loop}_{\mu\nu}}{\partial\varrho} = I^{\rm
ren}_{\mu\nu}(x)\sim \frac{G^2 m^2}{2c^2 r^2}
\left[\frac{\delta^0_{\mu} \delta^0_{\nu}}{c^2}(\varrho + 1) -
\eta_{\mu\nu} \varrho\right].
\end{eqnarray}
\noindent The right hand side of this equation cannot be
represented in the form (\ref{gaugesymeff}). This result can be
made more expressive by calculating the $\varrho$-variation of the
scalar curvature $R$ in a given point $\sigma^0$ of the reference
frame
\begin{eqnarray}
\delta R[h^{\rm eff}(x(\sigma^0))] &=& \frac{\delta R}{\delta
h^{\rm eff}_{\mu\nu}} \left(\frac{\partial h^{\rm
tree}_{\mu\nu}}{\partial\varrho} + \frac{\partial h^{\rm
loop}_{\mu\nu}}{\partial\varrho}\right)\delta\varrho +
\frac{\partial R}{\partial x^{\mu}}\delta x^{\mu}
\nonumber\\
&=& \frac{\partial R}{\partial x^{\alpha}}\Xi_{\rm tree}^{\alpha}
+ \left(\partial^{\mu}\partial^{\nu} - \eta^{\mu\nu}\Box\right)
\frac{h_{\mu\nu}^{\rm loop}}{\partial\varrho}\delta\varrho -
\frac{\partial R}{\partial x^{\mu}}\Xi_{\rm tree}^{\mu}
\nonumber\\
&=&
\partial_{i}\partial_{k}\frac{\partial h^{\rm loop}_{ik}}{\partial\varrho}
\delta\varrho + \Delta \frac{\partial h^{\rm
loop}}{\partial\varrho}\delta\varrho = \frac{G^2 m^2}{c^4r^4}(1 -
2\varrho)\delta\varrho\,, \nonumber
\end{eqnarray}
\noindent or\footnote{Another way to obtain this result is to
introduce the sources $t R$ and $k R_{,\alpha}c^{\alpha}$ for the
scalar curvature and its BRST-variation, respectively, into the
generating functional (\ref{gener}), instead of the corresponding
sources for the metric. Then Eq.~(\ref{hvarfa}) is replaced by
\begin{eqnarray}\label{hvarfr}
\frac{\partial R^{\rm eff}}{\partial\varrho} =
\left.\frac{\delta\langle\bar{c}^{\alpha}
\partial F_{\alpha}/\partial\varrho\rangle}
{\delta k}\right|_{\genfrac{}{}{0pt}{}{J\setminus j = 0}{K =
0}}\,.
\end{eqnarray}
\noindent At the second order in $G,$ the nontrivial contribution
comes again from the diagram of Fig.~\ref{fig4}(a) in which the
lower left vertex is now generated by $R_{,\alpha}c^{\alpha}.$
Thus, only the linear part of $R$ gives rise to a non-zero
contribution to the right hand side of Eq.~(\ref{hvarfr}). In
other words, $\delta R[h^{\rm eff}] = \delta R^{\rm eff}$, though
generally $R[h^{\rm eff}] \ne R^{\rm eff}.$}
\begin{eqnarray}\label{finalnew}
\frac{\partial R[h^{\rm eff}(x(\sigma^0))]}{\partial\varrho} =
\frac{G^2 m^2}{c^4r^4}(1 - 2\varrho)\,.
\end{eqnarray}
\noindent Equations (\ref{mainnew}), (\ref{finalnew}) express
violation of general covariance by the loop corrections.

{\it Thus, despite their independence of the Planck constant, the
post-Newtonian loop contributions turn out to be of a purely
quantum nature.}

We are now in a position to ask for conditions to be imposed on a
system in order to allow classical consideration of its
gravitational field. Such a condition providing vanishing of the
$\hbar^0$ loop contributions can easily be found out by examining
their dependence on the number of particles in the system. Let us
consider a body with mass $M,$ consisting of a large number $N =
M/m$ of elementary particles with mass $m.$ Then it is readily
seen that the $n$-loop contribution to the effective gravitational
field of the body turns out to be suppressed by a factor $1/N^n$
in comparison with the tree contribution. For instance, at the
first post-Newtonian order, the tree diagram in Fig.~\ref{fig1}(b)
is {\it bilinear} in the energy-momentum tensor $\langle
\hat{T}^{\mu\nu}\rangle$ of the particles, and therefore
proportional to $(m \cdot N) \cdot (m\cdot N) = M^2.$ On the other
hand, the post-Newtonian contribution of the diagram in
Fig.~\ref{fig2}(a) is proportional to $m^2 \cdot N = M^2/N,$ since
it has only two external matter lines.

Thus, we are led to the following macroscopic formulation of the
correspondence principle in quantum gravity: {\it the effective
gravitational field produced by a macroscopic body of mass $M$
consisting of $N$ particles turns into corresponding classical
solution of the Einstein equations in the limit $N \to \infty$}
\cite{kazakov1}. In particular, the principle of general
covariance is to be considered as approximate, valid only for the
description of macroscopic phenomena.

The $\hbar^0$-order loop contributions are normally highly
suppressed. For the solar gravitational field, for instance, their
relative value is $m_{\rm proton}/M_{\odot} \approx 10^{-57}.$
However, they are the larger the more gravitating body resembles
an elementary particle, and can become noticeable for a
sufficiently massive compact body.

\subsection{Effective gravitational field of a heavy particle}

As an application of the obtained results, we will calculate the
effective gravitational field of a particle with mass $M$ in the
first post-Newtonian approximation.

The complete expression of the order $\hbar^0$ for the spacetime
metric is the sum of two pieces. The first is the tree
contribution corresponding to the classical Schwarzschild solution
\begin{eqnarray}\label{sch1}
ds_{\rm cl}^2 = \left(1-\frac{r_g}{r}\right) c^2 d t^2 - \frac{d
r^2}{1-\frac{r_g}{r}} - r^2 (d\theta^2 + \sin^2\theta\
d\varphi^2),
\end{eqnarray}
\noindent where $\theta, \varphi$ are the standard spherical
angles, $r$ is the radial coordinate, and $r_g = 2 G M/c^2$ is the
gravitational radius of a spherically-symmetric distribution of
mass $M.$ The second is the one-loop post-Newtonian correction
contained in the diagram (a) of Fig.~\ref{fig2}. Restoring the
ordinary units, and using expressions (55), (65) of
Ref.~\cite{donoghue} for the vertex formfactors, and (\ref{hprop})
with $\varrho = 0 $ for the graviton propagator,\footnote{Note the
notation differences between Ref.~\cite{donoghue} and this
Chapter.} we obtain
\begin{eqnarray}&&\label{correctionp}
h^{{\rm loop}}_{\mu\nu}(p) = - \frac{\pi^2 G^2}{c^2 \sqrt{-p^2}}
\left(3 M^2 \eta_{\mu\nu} + \frac{q_{\mu} q_{\nu}}{c^2} + 7
M^2\frac{p_{\mu} p_{\nu}}{p^2}\right).
\end{eqnarray}
\noindent Written down in the coordinate space with the help of
the formulas (\ref{rootint}) and
\begin{eqnarray}
\int \frac{d^3{\bm{p}}}{(2\pi)^3}\frac{p_i
p_k}{|{\bm{p}}|^3}e^{i{\bm{p x}}} &=& \frac{1}{2\pi^2
r^2}\left(\delta_{i k} - \frac{2 x_i x_k }{r^2}\right)\,,
\nonumber
\end{eqnarray}
\noindent equation (\ref{correctionp}) gives, in the static case,
\begin{eqnarray}&&\label{correctionr}
h^{\rm loop}_{00} = - \frac{2 G^2 M^2}{c^2 r^2}\,, ~~h^{{\rm
loop}}_{i k} = \frac{G^2 M^2}{c^2 r^2}\left(- 2 \delta_{i k} +
\frac{7 x_i x_k}{r^2} \right).
\end{eqnarray}
\noindent Before adding the two contributions, however, one has to
transform Eq.~(\ref{sch1}) which form is fixed by the requirements
$g_{ti} = 0,$ $i = r, \theta, \varphi;$ $g_{\theta\theta} = -
r^2,$ to the DeWitt gauge
\begin{eqnarray}&&\label{dwgauge}
F_{\gamma} = \eta^{\mu\nu}\partial_{\mu}h_{\nu\gamma} -
\frac{1}{2}\partial_{\gamma}h\,, \qquad \zeta_{\alpha\beta} =
\eta_{\alpha\beta}\,,
\end{eqnarray}
\noindent under which Eq.~(\ref{correctionp}) was derived.
According to the general theorems about
$\zeta_{\alpha\beta}$-independence of the $\hbar^0$-order
contributions, proved in Sec.~\ref{feynman}, we can transform our
expressions to the singular case $\zeta_{\alpha\beta} = 0$ instead
of $\zeta_{\alpha\beta} = \eta_{\alpha\beta}.$ Then the effective
gravitational field will satisfy
\begin{eqnarray}\label{sgauge}
\eta^{\mu\nu}\partial_{\mu}h^{\rm eff}_{\nu\gamma} -
\frac{1}{2}\partial_{\gamma}h^{\rm eff} = 0\,.
\end{eqnarray}
\noindent The $t, \theta, \varphi$-components of
Eq.~(\ref{sgauge}) are already satisfied by the classical solution
(\ref{sch1}). To meet the remaining condition, let us substitute
$r \to f(r),$ where $f$ is a function of $r$ only. Then the $t,
\theta, \varphi$-components of Eq.~(\ref{gauge}) are still
satisfied, while its $r$-component gives the following equation
for the function $f(r):$
\begin{eqnarray}\label{rcond}
\frac{1}{r^2}\frac{\partial}{\partial r}\left(\frac{r^2
f'^2}{1-r_g/f}\right) -\frac{2 f^2}{r^2} -
\frac{1}{2}\frac{\partial}{\partial r}\left(1-\frac{r_g}{f} +
\frac{f'^2}{1-r_g/f}+\frac{2 f^2}{r^2}\right) = 0,
\end{eqnarray}
\noindent where $f'\equiv \partial f(r)/\partial r.$

Since we are interested only in the long-distance corrections to
the Newton law, one may expand $f(r)/r$ in powers of $r_g/r$
keeping only the first few terms:
$$f(r) = r\left[1 + c_1 \frac{r_g}{r} + c_2 \left(\frac{r_g}{r}\right)^2
+ \cdot\cdot\cdot \right].$$ Substituting this into
Eq.~(\ref{rcond}), one obtains successively $c_1 =1/2,$ $c_2=1/2,$
etc. Therefore, up to terms of the order $r^2_g/r^2,$ the
Schwarzschild solution takes the following form
\begin{eqnarray}\label{sch2}&&
ds_{\rm cl}^2 = \left(1 - \frac{r_g}{r} + \frac{r^2_g}{2
r^2}\right) c^2 d t^2 - \left(1 + \frac{r_g}{r} - \frac{r^2_g}{2
r^2}\right) d r^2 \nonumber\\&& - r^2 \left(1 + \frac{r_g}{r} +
\frac{5 r^2_g}{4 r^2}\right) (d\theta^2 + \sin^2\theta\
d\varphi^2)\,.
\end{eqnarray}\noindent
Rewriting Eq.~(\ref{correctionr}) in spherical coordinates and
adding the tree contribution, we finally obtain the following
expression for the interval \cite{kazakov3}
\begin{eqnarray}\label{eff}&&
ds_{\rm eff}^2 \equiv g^{\rm eff}_{\mu\nu} d x^{\mu} d x^{\nu} =
\left(1 - \frac{r_g}{r}\right) c^2 d t^2 - \left(1 + \frac{r_g}{r}
- \frac{7 r^2_g}{4 r^2}\right) d r^2 \nonumber\\&& - r^2 \left(1 +
\frac{r_g}{r} + \frac{7 r^2_g}{4 r^2}\right) (d\theta^2 +
\sin^2\theta\ d\varphi^2)\,.
\end{eqnarray}\noindent
In connection with this result it should be noted the following.
Taking into account higher-order radiative post-Newtonian
corrections will result in a further modification of the
Schwarzschild solution. Since quantum contributions are of the
same order of magnitude as those given by general relativity, this
modification can lead to a significant shift of the horizon. In
particular, the metric $g^{\rm eff}_{\mu\nu}(r)$ may well turn out
to be a regular function of $r$ when all the $\hbar^0$ loop
corrections are taken into account.

\section{Quantum fluctuations of gravitational field}\label{qflucs}

We turn now to another aspect of the long-range behavior in
quantum gravity -- quantum fluctuations of the gravitational
fields produced by elementary systems. On various occasions, the
issue of fluctuations has been the subject of a number of
investigations (see Refs.~\cite{zerbini,ford2,ford3,hu1,hu2},
where references to early works can be found). It should be
mentioned, however, that despite extensive literature in the area,
only vacua contributions of quantized matter fields to the metric
fluctuations have been studied in detail. At the same time, it is
effects produced by real matter that are of special interest
concerning the structure of elementary contributions to the field
fluctuation.

Before we proceed to calculation of the correlation function, we
shall examine its general properties in more detail. Namely, the
structure of the long-range expansion of the correlation function,
and the question of its gauge dependence will be considered.

\subsection{Properties of correlation
function}\label{cfproperties}

As defined by Eq.~(\ref{green2}), the correlation function
$C_{\mu\nu\alpha\beta}$ is a function of two spacetime arguments
$(x, x').$ Of special interest is its ``diagonal element''
corresponding to coinciding arguments and Lorentz indices $(x=x',
\mu\nu = \alpha\beta),$ which describes a dispersion of the
spacetime metric around its mean value in a given spacetime point.
However, it is well-known that this element is not generally
well-defined because of the singular behavior of the product of
field operators in the coincidence limit $x \to x'.$ This
difficulty is naturally resolved when one takes into account the
fact (realized long ago, see, e.g., Ref.~\cite{bohr1}) that in any
field measurement in a given spacetime point, one deals actually
with the field averaged over a small but finite spacetime domain
surrounding this point. Thus, the physically sensible expression
for the field operator is the following
\begin{eqnarray}\label{stav}
\hat{\EuScript{H}}_{\mu\nu} = \frac{1}{VT}\int_T dt\int_V d^3
\bm{x}~\hat{h}_{\mu\nu}(\bm{x},t)\,.
\end{eqnarray}
\noindent Respectively, the product of two fields in a given point
is understood as the limit of
\begin{eqnarray}\label{stavprod}
\hat{\EuScript{B}}_{\mu\nu\alpha\beta} = \frac{1}{(VT)^2}\int_T
dt\int_T dt'\int_V d^3\bm{x} \int_V d^3
\bm{x}'~\hat{h}_{\mu\nu}(\bm{x},t)
\hat{h}_{\alpha\beta}(\bm{x}',t')
\end{eqnarray} \noindent when the size of the domain tends to zero.
Finally, correlation function of the spacetime metric in this
domain is
\begin{eqnarray}\label{a0av}
\EuScript{C}_{\mu\nu\alpha\beta} = \langle
\hat{\EuScript{B}}_{\mu\nu\alpha\beta}\rangle -
\langle\hat{\EuScript{H}}_{\mu\nu}\rangle
\langle\hat{\EuScript{H}}_{\alpha\beta}\rangle \equiv
\frac{1}{(VT)^2}\iint d^4 x d^4 x' C_{\mu\nu\alpha\beta}(x,x')\,.
\end{eqnarray}
\noindent Figure \ref{fig6} depicts the tree diagrams contributing
to the right hand side of Eq.~(\ref{fintctp1}). The disconnected
part shown in Fig.~\ref{fig6}(a) cancels in the expression for the
correlation function, Eq.~(\ref{a0av}), which is thus represented
by the diagrams (b)--(h).

\subsubsection{Correlation function in the long-range
limit}\label{lr}

As we mentioned in the beginning of Sec.~\ref{transformegf}, the
mean gravitational field produced by a massive particle is a
function of five dimensional parameters -- the fundamental
constants $\hbar,G,c,$ the particle's mass $m,$ and the distance
between the particle and the point of observation, $r.$ Unlike the
mean field, however, $\EuScript{C}_{\mu\nu\alpha\beta}$ is a
function of two spacetime arguments. On the other hand, we are
interested ultimately in the coincidence limit of this function,
which will be shown below to exist everywhere except the region of
particle localization. In this limit, therefore,
$\EuScript{C}_{\mu\nu\alpha\beta}$ depends on the same five
parameters $\hbar,G,c,m,r.$ Assuming as before the particle
sufficiently heavy ($\varkappa \to 0$), we conclude that the
relevant information about field correlations is contained in the
long-range asymptotic of $\EuScript{C}_{\mu\nu\alpha\beta}(r).$

To extract this information we note, first of all, that in the
long-range limit, the value of $\EuScript{C}_{\mu\nu\alpha\beta}$
is independent of the choice of spacetime domain used in the
definition of physical gravitational field operators,
Eq.~(\ref{stav}). Indeed, in any case the size of this domain must
be small in comparison with the characteristic length at which the
mean field changes significantly. In the case considered, this
requires that $V \ll r^3.$ To the leading order of the long-range
expansion, therefore, the quantity $\langle\hat{h}_{\mu\nu}(x)
\hat{h}_{\alpha\beta}(x')\rangle$ appearing in the right hand side
of Eq.~(\ref{a0av}) can be considered constant within the domain.
However, one cannot set $x=x'$ in this expression directly. It is
not difficult to see that the formal expression
$\langle\hat{h}_{\mu\nu}(x) \hat{h}_{\alpha\beta}(x)\rangle$ does
not exist. Consider, for instance, the diagram shown in
Fig.~\ref{fig6}(b). A typical term in the analytical expression of
this diagram is proportional to the integral
\begin{eqnarray}&&
I_{\mu\nu\alpha\beta}(x-x',p) = \int d^4 k
\frac{V_{\mu\nu\alpha\beta}\theta(k^0)\delta(k^2)e^{ik(x-x')}}{[(k+q)^2
- m^2](k-p)^2}\ ,
\end{eqnarray}
where $q_{\mu}$ is the 4-momentum of the particle, $p_{\mu}$ the
momentum transfer, and $V_{\mu\nu\alpha\beta}$ a vertex factor
combined of $\eta_{\mu\nu}$'s and momenta $q,p,k.$ For a small but
nonzero $(x-x')$ this integral is effectively cut-off at large
$k$'s by the oscillating exponent, but for $x=x'$ it is divergent.
This divergence arises from integration over large values of
virtual graviton momenta, and therefore has nothing to do with the
long-range behavior of the correlation function, because this
behavior is determined by the low-energy properties of the theory.
Evidently, the singularity of $I_{\mu\nu\alpha\beta}(x-x',p)$ for
$x \to x'$ is not worse than $\ln(x^{\mu} -
x^{\prime\mu})^{2}/(x^{\mu} - x^{\prime\mu})^{2}.$ Therefore,
$\langle {\rm in}
|\hat{h}_{\mu\nu}(x)\hat{h}_{\alpha\beta}(x')|{\rm in}\rangle$ is
integrable, and $\EuScript{C}_{\mu\nu\alpha\beta}$ given by
Eqs.~(\ref{stavprod}), (\ref{a0av}) is well defined.

Our aim below will be to show that this singularity can be
consistently isolated and removed from the expression for
$I_{\mu\nu\alpha\beta}(x-x',p)$ and similar integrals for the rest
of diagrams in Fig.~\ref{fig6}, without changing  the long-range
properties of the correlation function. After this removal, it is
safe to set $x=x'$ in the finite remainder, and to consider
$\EuScript{C}_{\mu\nu\alpha\beta}$ as a function of the single
variable -- the distance $r.$ An essential point of this procedure
is that the singularity turns out to be {\it local}, and hence
does not interfere with terms describing the long-range behavior,
which guaranties unambiguity of the whole procedure.

It should be emphasized that in contrast to what takes place in
the scattering theory, the ultraviolet divergences appearing in
the course of calculation of the in-in matrix elements in the
coincidence limit are generally non-polynomial with respect to the
momentum transfer. The reason for this is the different analytic
structure of various elements in the matrix propagators
$\mathfrak{D}_{\mu\nu\alpha\beta},$ $\mathfrak{D},$ which spoils
the simple ultraviolet properties exhibited by the ordinary
Feynman amplitudes.\footnote{As is well known, the proof of
locality of the S-matrix divergences relies substantially on the
causality of the pole structure of Feynman propagators. This
property allows Wick rotation of the energy contours, thus
revealing the essentially Euclidean nature of the ultraviolet
divergences.} Take the above integral as an example. Because of
the delta function in the integrand, differentiation of
$I_{\mu\nu\alpha\beta}(x-x',p)$ with respect to the momentum
transfer does not remove the ultraviolet divergence of
$I_{\mu\nu\alpha\beta}(0,p).$ What makes it all the more
interesting is the result obtained in Sec.~\ref{calcul} below,
that the non-polynomial parts of divergent contributions
eventually cancel each other, and the overall divergence turns out
to be completely local.

Next, let us establish general form of the leading term in the
long-range expansion of the correlation function. In momentum
representation, an expression of lowest order in the momentum
transfer with suitable dimensionality is the following
\begin{eqnarray}\label{zc}
\EuScript{C}^{(0)}_{\mu\nu\alpha\beta}(p) \sim G^2 m^2
\frac{\EuScript{F}_{\mu\nu\alpha\beta}}{\sqrt{-p^2}}\ ,
\end{eqnarray}
\noindent where $\EuScript{F}_{\mu\nu\alpha\beta}$ is a
dimensionless positive definite tensor combined from
$\eta_{\mu\nu}$ and $q_{\mu}.$

Not all of diagrams in Fig.~\ref{fig6} contain contributions of
this type. It is not difficult to identify those which do not.
Consider, for instance, the diagram (h). It is proportional to the
integral
$$\int d^4 k
\theta(k^0)\delta(k^2)\frac{e^{ik(x-x')}}{(k-p)^2}\ ,$$ which does
not involve the particle mass at all. Taking into account $m^2$
coming from the vertex factor, and $(2\varepsilon_{\bm{q}})^{-1/2}
\sim \sqrt{m}$ from each external matter line, we see that the
contribution of the diagram (h) is proportional to $m,$ not $m^2.$
The same is true of all other diagrams without internal matter
lines. As to diagrams involving such lines, it will be shown in
Sec.~\ref{calcul} by direct calculation that they do contain
contributions of the type Eq.~(\ref{zc}). But prior to this the
question of their dependence on the gauge will be considered.

\subsubsection{Gauge independence of the leading
contribution}\label{gd}

As in the case of mean gravitational field considered in
Sec.~\ref{transformegf}, the 0-component of the momentum transfer
is to be set zero when calculating the leading term in the
long-range expansion of the correlation function. This implies, in
particular, that $\EuScript{C}^{(0)}_{\mu\nu\alpha\beta}$ contains
information about fluctuations in a quantity of direct physical
meaning -- the static potential energy of interacting particles.
As such, it is expected to be independent of the choice of gauge
conditions used to fix general covariance. More precisely, we have
to verify that under variations of the gauge conditions,
$\EuScript{C}^{(0)}_{\mu\nu\alpha\beta}$ varies in a way that does
not affect the values of observables built from it.

As we saw in Sec.~\ref{lr}, the only diagrams contributing in the
long-range limit are those containing internal massive particles
lines. We will show presently that the gauge-dependent part of
these diagrams can be reduced to the form without such lines.
First, it follows form Eq.~(\ref{aprop}) that the gauge variation
of the graviton propagator satisfies
\begin{eqnarray}\label{xideriv}
\delta\mathfrak{D}_{\mu\nu\sigma\lambda} =
\mathfrak{D}_{\mu\nu\alpha\beta}
\delta\mathfrak{G}^{\alpha\beta\gamma\delta}
\mathfrak{D}_{\gamma\delta\sigma\lambda}\,.
\end{eqnarray}
\noindent On the other hand, contracting Eq.~(\ref{aprop}) with
$G^{(0)\rho}_{\mu\nu}$ yields
$$\mathfrak{i}\,F_{\alpha}^{,\mu\nu}
G^{(0)\rho}_{\mu\nu}\xi^{\alpha\beta} F^{,\gamma\delta}_{\beta}
\mathfrak{D}_{\gamma\delta\sigma\lambda} = - \mathfrak{e}
G^{(0)\rho}_{\sigma\lambda}.$$ Defining the matrix ghost
propagator $\tilde{\mathfrak{D}}_{\beta}^{\alpha}$ according to
Eqs.~(\ref{aprop}), (\ref{ghostp}) by
\begin{eqnarray}&&
\mathfrak{i}F_{\alpha}^{,\mu\nu}G^{(0)\beta}_{\mu\nu}
\tilde{\mathfrak{D}}^{\gamma}_{\beta} = - \mathfrak{e}
\delta_{\alpha}^{\gamma}\, \nonumber
\end{eqnarray}
\noindent one finds
\begin{eqnarray}\label{slavnew}
F_{\alpha}^{,\mu\nu} \mathfrak{D}_{\mu\nu\sigma\lambda} =
\zeta_{\alpha\gamma}
G^{(0)\beta}_{\sigma\lambda}\tilde{\mathfrak{D}}_{\beta}^{\gamma}\,,
\end{eqnarray}
\noindent Taking into account that
$$\delta\mathfrak{G}^{\alpha\beta\gamma\delta} = \mathfrak{i}
\delta
F^{,\alpha\beta}_{\rho}\xi^{\rho\tau}F^{,\gamma\delta}_{\tau} +
\mathfrak{i}F^{,\alpha\beta}_{\rho}\xi^{\rho\tau}\delta
F^{,\gamma\delta}_{\tau}\,,$$ and using Eq.~(\ref{slavnew}), we
get
\begin{eqnarray}\label{deltad}
\delta\mathfrak{D}_{\mu\nu\sigma\lambda} = \left(\delta
F_{\delta}^{,\alpha\beta}
\mathfrak{D}_{\mu\nu\alpha\beta}\,\mathfrak{i}\,
\tilde{\mathfrak{D}}_{\gamma}^{\delta}\right)
G^{(0)\gamma}_{\sigma\lambda} +
G^{(0)\beta}_{\mu\nu}\left(\tilde{\mathfrak{D}}_{\beta}^{\alpha}
\,\mathfrak{i}\,\delta F_{\alpha}^{,\gamma\delta}
\mathfrak{D}_{\gamma\delta\sigma\lambda} \right) \,.
\end{eqnarray}
\noindent Let us assume for definiteness that the pair of indices
$\mu\nu$ refers to the point of observation, and consider the
first term in Eq.~(\ref{deltad}). This part of the gauge variation
of the graviton propagator is attached to the matter line through
the generator $G^{(0)\gamma}_{\sigma\lambda}.$ Suppressing all
Lorentz and matrix indices except those referring to the $\phi^2
h$-vertex, it can be written as
$$\mathfrak{b}_{\sigma\lambda} =
\left(\begin{array}{c} a_{\gamma+}  \\
a_{\gamma-}
\end{array}\right)G^{(0)\gamma}_{\sigma\lambda}\,,
\qquad \mathfrak{a}_{\gamma} = \delta F_{\delta}^{,\alpha\beta}
\mathfrak{D}_{\mu\nu\alpha\beta}\,\mathfrak{i}\,
\tilde{\mathfrak{D}}_{\gamma}^{\delta}\,.$$ The matrix vertex
$\mathfrak{V}^{\sigma\lambda,ik}$ is obtained multiplying
$\delta^3 S_\phi /\delta\phi_i\delta\phi_k\delta
h_{\sigma\lambda}$ by the matrix $e_{stu}.$
Equation~(\ref{ident1}) then shows that the combination
$\mathfrak{b}_{\sigma\lambda}\mathfrak{V}^{\sigma\lambda,ik}$ can
be written as
\begin{eqnarray}\label{ident2}
\mathfrak{b}_{\sigma\lambda}\mathfrak{V}^{\sigma\lambda,ik} &=& -
\mathfrak{f}_{\gamma}\left\{\frac{\delta
S^{(2)}_{\phi}}{\delta\phi_l \delta\phi_k}\frac{\delta
G_l^{\gamma}}{\delta\phi_i} + \frac{\delta
S^{(2)}_{\phi}}{\delta\phi_l \delta\phi_i}\frac{\delta
G_l^{\gamma}}{\delta\phi_k}\right\}, \nonumber
\end{eqnarray}
\noindent  where $$\mathfrak{f}_{\gamma} = \left(\begin{array}{cc}
a_{\gamma+}&0\\0&-a_{\gamma-}\end{array}\right)\,.$$ Finally,
contracting
$\mathfrak{b}_{\sigma\lambda}\mathfrak{V}^{\sigma\lambda,ik}$ with
the matrix $\mathfrak{D}_{km}$ and the vector
$\tilde{\mathfrak{r}}_i =
\left(\bar{\phi}_{i},\bar{\phi}_{i}\right)\,,$ and using
Eqs.~(\ref{aprop1}), (\ref{free}) gives
\begin{eqnarray}
\tilde{\mathfrak{r}}_i\mathfrak{b}_{\sigma\lambda}
\mathfrak{V}^{\sigma\lambda,ik}\mathfrak{D}_{km} =
\tilde{\mathfrak{a}}_{\gamma}\frac{\delta
G_m^{\gamma}}{\delta\phi_i}\bar{\phi}_{i}\,. \nonumber
\end{eqnarray}
\noindent Similarly to what we have found in Sec.~\ref{feynman}
considering $\zeta$-dependence of the one-loop contribution to the
effective metric, the matrix matter propagator is cancelled upon
contraction with the vertex factor. As in the case of the one-loop
post-Newtonian contributions, the $\hbar^0$ terms in the
correlation function are associated with the virtual matter quanta
propagating near their mass shells. Hence, the first term in the
gauge variation of the propagator, Eq.~(\ref{deltad}), results to
terms of higher order in the Planck constant. As to the second
term, it gives rise to a $\hbar^0$ order variation of the
correlation function, but its structure with respect to the
indices $\mu\nu$ is that of an ordinary gauge transformation, so
$$\delta\EuScript{C}^{(0)}_{\mu\nu\alpha\beta}
\sim G^{\gamma}_{\mu\nu}\Omega_{\gamma,\alpha\beta} +
G^{\gamma}_{\alpha\beta}\Omega_{\gamma,\mu\nu}$$ with some
infinitesimal $\Omega_{\gamma,\alpha\beta},$ which proves gauge
independence of gauge-invariant functionals built from
$\EuScript{C}^{(0)}_{\mu\nu\alpha\beta}.$ Note also that as far as
variations of the weighting matrix $\xi^{\alpha\beta}$ are
considered, not only these functionals, but also
$\EuScript{C}^{(0)}_{\mu\nu\alpha\beta}$ itself turns out to be
gauge independent. Indeed, a variation $\delta\xi^{\alpha\beta}$
of the weighting matrix is equivalent to the variation of the
gauge condition $F_{\alpha}:$
$$\delta F_{\alpha} = \theta^{\beta}_{\alpha}F_{\beta}\,,$$
provided that $$\delta\xi^{\alpha\beta} =
\xi^{\alpha\gamma}\theta^{\beta}_{\gamma} +
\xi^{\beta\gamma}\theta^{\alpha}_{\gamma}\,.$$ Then using
Eq.~(\ref{slavnew}) in Eq.~(\ref{deltad}) gives
$$\delta\mathfrak{D}_{\mu\nu\sigma\lambda} = - G^{(0)\beta}_{\mu\nu}
\tilde{\mathfrak{D}}_{\beta}^{\alpha}
\,\mathfrak{i}\,\delta\zeta_{\alpha\delta}
\tilde{\mathfrak{D}}_{\gamma}^{\delta}
G^{(0)\gamma}_{\sigma\lambda}\,,$$ which implies that
$\EuScript{C}^{(0)}_{\mu\nu\alpha\beta}$ is independent of the
choice of the matrix $\zeta_{\alpha\beta}.$

\subsection{Evaluation of the leading contribution}
\label{calcul}

\begin{figure}
\includegraphics{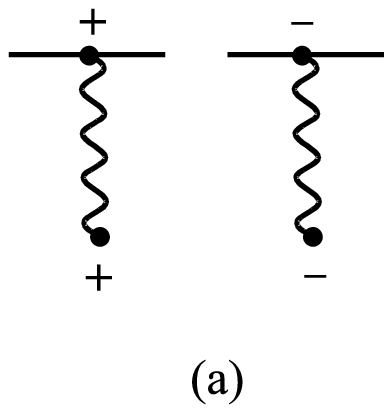}
\includegraphics{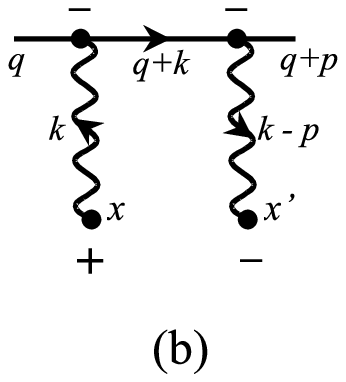}
\includegraphics{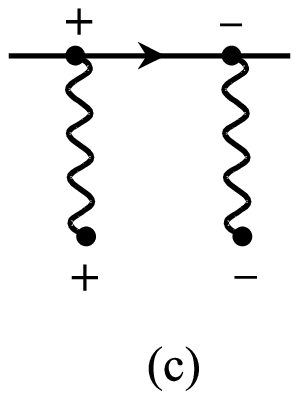}
\includegraphics{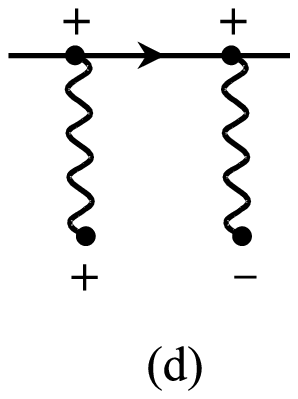}
\includegraphics{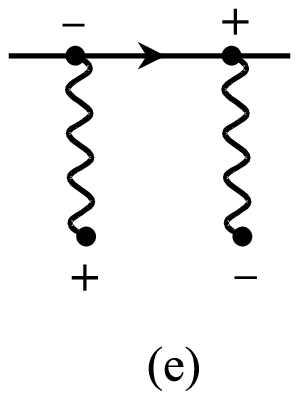}
\includegraphics{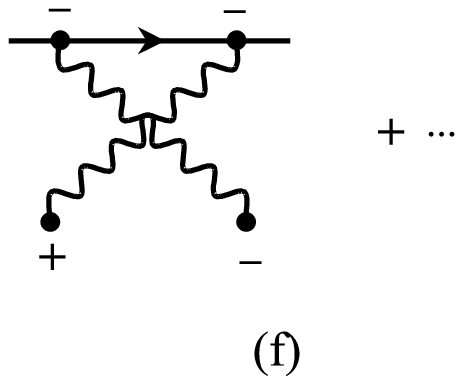}\\
\includegraphics{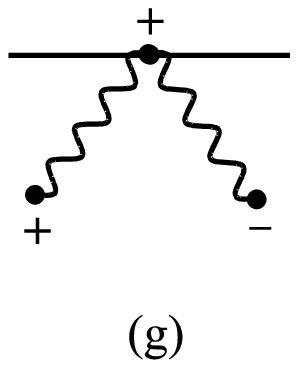}
\includegraphics{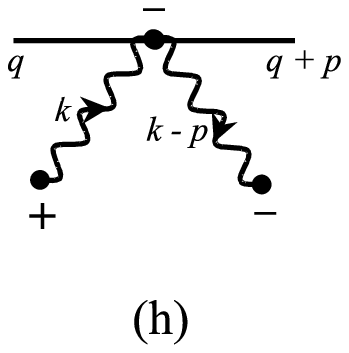} \caption{Tree
contribution to the right hand side of Eq.~(\ref{fintctp1}). Part
(f) of the figure represents the ``transposition'' of diagrams
(b)--(e) (see Sec.~\ref{calcul}).} \label{fig6}
\end{figure}

Let us proceed to the calculation of the leading contribution to
the correlation function, assuming as in Sec.~\ref{general} that
the field producing particle is a scalar described by the action
(\ref{actionm}). This contribution is contained in the sum of
diagrams (b)--(f) in Fig.~\ref{fig6}, which has the symbolic form
$$C_{\mu\nu\alpha\beta} = I_{\mu\nu\alpha\beta}
+ I^{\rm tr}_{\mu\nu\alpha\beta}\,, \qquad I_{\mu\nu\alpha\beta} =
\frac{1}{i}\left\{\mathfrak{D}_{\mu\nu\sigma\lambda}\left[\mathfrak{r}^\dag
\mathfrak{V}^{\sigma\lambda}\mathfrak{D}
\mathfrak{V}^{\tau\rho}\mathfrak{r}\right]
\mathfrak{D}_{\tau\rho\alpha\beta}\right\}_{+-}\,,$$ where the
superscript ``tr'' means transposition of the indices and
spacetime arguments referring to the points of observation:
$\mu\nu\leftrightarrow\alpha\beta,$ $+\leftrightarrow -,$
$x\leftrightarrow x'$ [the transposed contribution is represented
by the diagrams collected in part (f) of Fig.~\ref{fig6}].

As it follows from the considerations of Sec.~\ref{lr},
$\EuScript{C}^{(0)}_{\mu\nu\alpha\beta}$ can be expressed through
$C_{\mu\nu\alpha\beta}$ as
\begin{eqnarray}\label{c0i0}
\EuScript{C}^{(0)}_{\mu\nu\alpha\beta} =
\lim\limits_{\genfrac{}{}{0pt}{}{m\to\infty}{V,T\to 0
}}\left\{\frac{1}{(VT)^2}\iint\limits_{(V,T)} d^4 x d^4 x'~
C_{\mu\nu\alpha\beta}(x,x')\right\}\,.
\end{eqnarray}
\noindent In the DeWitt gauge
$$\mathfrak{D}_{\mu\nu\alpha\beta} = -
W_{\mu\nu\alpha\beta}\mathfrak{D}^{0}\,, \qquad \mathfrak{D}^{0}
\equiv \mathfrak{D}|_{m=0}\,,$$ $I_{\mu\nu\alpha\beta}$ reads
\begin{eqnarray}\label{diagen}
I_{\mu\nu\alpha\beta}(x,x') &=& \frac{1}{i}\iint d^4 z d^4 z'\nonumber\\
\times \Biggl\{&+& D^0_{++}(x,z)\left[\bar{\phi}(z)
\stackrel{\leftrightarrow}{V}_{\mu\nu}
D_{++}(z,z')\stackrel{\leftrightarrow}{V'}_{\alpha\beta}
\bar{\phi}(z')\right]D^0_{+-}(z',x') \nonumber\\ &-&
D^0_{++}(x,z)\left[\bar{\phi}(z)
\stackrel{\leftrightarrow}{V}_{\mu\nu}
D_{+-}(z,z')\stackrel{\leftrightarrow}{V'}_{\alpha\beta}\bar{\phi}(z')\right]D^0_{--}(z',x')\nonumber\\
&-& D^0_{+-}(x,z)\left[\bar{\phi}(z)
\stackrel{\leftrightarrow}{V}_{\mu\nu}
D_{-+}(z,z')\stackrel{\leftrightarrow}{V'}_{\alpha\beta}
\bar{\phi}(z')\right]D^0_{+-}(z',x')\nonumber\\ &+&
D^0_{+-}(x,z)\left[\bar{\phi}(z)
\stackrel{\leftrightarrow}{V}_{\mu\nu}
D_{--}(z,z')\stackrel{\leftrightarrow}{V'}_{\alpha\beta}
\bar{\phi}(z')\right]D^0_{--}(z',x') \Biggr\}\,,\nonumber\\
\end{eqnarray}
\noindent where
$$\varphi\stackrel{\leftrightarrow}{V}_{\mu\nu}\psi =
\frac{1}{2}W_{\mu\nu\alpha\beta}
\left\{W^{\alpha\beta\gamma\delta}\varphi\stackrel{\leftarrow}{
\partial_{\gamma}}\stackrel{\rightarrow}
{\partial_{\delta}}\psi - m^2\eta^{\alpha\beta}\varphi\psi\right\}
\,.$$ Contribution of the third term in the right hand side of
Eq.~(\ref{diagen}) is zero identically. Indeed, using
Eq.~(\ref{explprop}), and performing spacetime integrations we see
that the three lines coming, say, into $z$-vertex are all on the
mass shell, which is inconsistent with the momentum conservation
in the vertex. The remaining terms in Eq.~(\ref{diagen}) take the
form
\begin{eqnarray}\label{diagenk1}
I_{\mu\nu\alpha\beta}(x,x') &=& \iint \frac{d^3 \bm{q}}{(2\pi)^3}
\frac{d^3 \bm{p}}{(2\pi)^3}\frac{a(\bm{q})a^*(\bm{q} +
\bm{p})}{\sqrt{2\varepsilon_{\bm q}2\varepsilon_{{\bm q} +
\bm{p}}}} e^{ipx'}\tilde{I}_{\mu\nu\alpha\beta}(p,q)\,, \quad p_0
= \varepsilon_{\bm{q} + \bm{p}} - \varepsilon_{\bm{q}}\,,
\end{eqnarray}
\noindent where
\begin{eqnarray}\label{diagenk2}
\tilde{I}_{\mu\nu\alpha\beta}(p,q) = &-& i\int \frac{d^4
k}{(2\pi)^4} e^{ik(x-x')}\{ m^4 \eta_{\mu\nu}\eta_{\alpha\beta} -
2 m^2 (\eta_{\mu\nu}q_{\alpha} q_{\beta} +
\eta_{\alpha\beta}q_{\mu} q_{\nu}) - 4 m^2\eta_{\mu\nu}q_{(\alpha}
k_{\beta)}
\nonumber\\
&-& 2 m^2 \eta_{\alpha\beta}(p_{(\mu} q_{\nu)} + p_{(\mu} k_{\nu)}
+ q_{(\mu} k_{\nu)}) + 4 p_{(\mu} q_{\nu)} q_{(\alpha} k_{\beta)}
\nonumber\\
&+& 4 q_{\alpha} q_{\beta} (p_{(\mu} q_{\nu)} + p_{(\mu} k_{\nu)})
+ 4 p_{(\mu} k_{\nu)} q_{(\alpha} k_{\beta)}
\nonumber\\
&+& 8 q_{(\mu} q_{\nu} q_{\alpha} k_{\beta)} + 4 q_{(\mu} k_{\nu)}
q_{(\alpha} k_{\beta)} + 4 q_{\mu} q_{\nu} q_{\alpha} q_{\beta}
\}\nonumber\\
&\times&\Bigl\{D^0_{++}(k)D_{++}(q+k)D^0_{+-}(k-p) \nonumber\\
&-&D^0_{++}(k)D_{+-}(q+k)D^0_{--}(k-p)\nonumber\\
&+&D^0_{+-}(k)D_{--}(q+k)D^0_{--}(k-p)\Bigr\}\,,
\end{eqnarray}
\noindent $(\mu_1\mu_2\cdots\mu_n)$ denoting symmetrization over
indices enclosed in the parentheses,
$$(\mu_1\mu_2\cdots\mu_n) = \frac{1}{n!}
\sum\limits_{\genfrac{}{}{0pt}{}{\{i_1 i_2\cdots i_n\} =} {{\rm
perm}\{ 12\cdots n\}}} \mu_{i_1}\mu_{i_2}\cdots\mu_{i_n}\,.$$

As in Sec.~\ref{general}, Eq.~(\ref{diagenk1}) simplifies in the
long-range limit to
\begin{eqnarray}\label{diagenk3}
I_{\mu\nu\alpha\beta}(x,x') &=& \frac{1}{2m}\iint\frac{d^3
\bm{q}}{(2\pi)^3}\frac{d^3 \bm{p}}{(2\pi)^3}|b(\bm{q})|^2
e^{-i\bm{p}(\bm{x'} - \bm{x}_0
)}\tilde{I}_{\mu\nu\alpha\beta}(p,q)\,, \quad p_0 = 0\,.
\end{eqnarray}
\noindent The leading term in $\tilde{I}_{\mu\nu}(p,q)$ has the
form [Cf. Eq.~(\ref{zc})]
\begin{eqnarray}\label{leading}
\tilde{I}^{(0)}_{\mu\nu\alpha\beta}(p,q) \sim G^2 m^2
\frac{\EuScript{F}_{\mu\nu\alpha\beta}}{\sqrt{-p^2}}\ .
\end{eqnarray}
\noindent This singular at $p\to 0$ contribution comes from
integration over small $k$ in Eq.~(\ref{diagenk2}). Therefore, to
the leading order, the momenta $k,p$ in the vertex factors can be
neglected in comparison with $q.$ Thus,
\begin{eqnarray}\label{diagenk4}
\tilde{I}_{\mu\nu\alpha\beta}(p,q) = -
iQ_{\mu\nu}Q_{\alpha\beta}\int\frac{d^4 k}{(2\pi)^4} e^{ik(x-x')}
\Bigl\{&&D^0_{++}(k)D_{++}(q+k)D^0_{+-}(k-p) \nonumber\\
&-&D^0_{++}(k)D_{+-}(q+k)D^0_{--}(k-p)\nonumber\\
&+&D^0_{+-}(k)D_{--}(q+k)D^0_{--}(k-p)\Bigr\}\,,
\end{eqnarray}
\noindent where $$Q_{\mu\nu} \equiv 2 q_{\mu}q_{\nu} -
m^2\eta_{\mu\nu}\,.$$ Furthermore, it is convenient to combine
various terms in this expression with the corresponding terms in
the transposed contribution. Noting that the right hand side of
Eq.~(\ref{diagenk4}) is explicitly symmetric with respect to
$\{\mu\nu\}\leftrightarrow\{\alpha\beta\}$ and that the variables
$x,x'$ in the exponent can be freely interchanged because they
appear symmetrically in Eq.~(\ref{c0i0}), we may write
\begin{eqnarray}\label{diagenk5}
\tilde{C}_{\mu\nu\alpha\beta}(p,q) &\equiv&
\tilde{I}_{\mu\nu\alpha\beta}(p,q) + \tilde{I}^{\rm
tr}_{\mu\nu\alpha\beta}(p,q) = - iQ_{\mu\nu}Q_{\alpha\beta}\int
\frac{d^4 k}{(2\pi)^4} e^{ik(x-x')}\nonumber\\
\times\Bigl\{&\;&\left[D^0_{++}(k)D_{++}(q+k)D^0_{+-}(k-p)
+ D^0_{--}(k)D_{--}(q+k)D^0_{-+}(k-p)\right]\nonumber\\
&+& \left[D^0_{+-}(k)D_{--}(q+k)D^0_{--}(k-p) +
D^0_{-+}(k)D_{++}(q+k)D^0_{++}(k-p)
\right]\nonumber\\
&-& \left[D^0_{++}(k)D_{+-}(q+k)D^0_{--}(k-p) +
D^0_{--}(k)D_{-+}(q+k)D^0_{++}(k-p)\right]\Bigr\}\,.
\end{eqnarray}
\noindent With the help of the relation
\begin{eqnarray}\label{aux2}
D_{--} + D_{++} = D_{+-} + D_{-+}\,,
\end{eqnarray}
\noindent which is a consequence of the identity
$$T\hat{\phi}(x)\hat{\phi}(y) + \tilde{T}\hat{\phi}(x)\hat{\phi}(y)
- \hat{\phi}(x)\hat{\phi}(y) - \hat{\phi}(y)\hat{\phi}(x) = 0\,,$$
the first term in the integrand can be transformed as
\begin{eqnarray}&&
D^0_{++}(k)D_{++}(q+k)D^0_{+-}(k-p) +
D^0_{--}(k)D_{--}(q+k)D^0_{-+}(k-p) \nonumber\\&& =
D^0_{++}(k)D_{++}(q+k)D^0(k-p) +
D^0(k)D_{--}(q+k)D^0_{-+}(k-p)\,,\nonumber
\end{eqnarray}
\noindent where $$D(k) \equiv D_{+-}(k) + D_{-+}(k) = 2\pi i
\delta(k^2 - m^2)\,, \quad D^0(k) = D(k)|_{m=0}\,.$$ Here we used
the already mentioned fact that $D^0_{+-}(k)D_{-+}(k+q) \equiv 0$
for $q$ on the mass shell. Analogously, the second term becomes
$$D^0(k)D_{--}(q+k)D^0_{--}(k-p) - D^0_{-+}(k)D_{--}(q+k)D^0(k-p)\,.$$
Changing the integration variables $k \to k + p,$ $\bm{q} \to
\bm{q} - \bm{p},$ and then $\bm{p} \to - \bm{p},$ noting that the
leading term is even in the momentum transfer [see
Eq.~(\ref{leading})], and that $\bm{p}(\bm{x}-\bm{x}')$ in the
exponent can be omitted in the coincidence limit ($x\to x'$), the
sum of the two terms takes the form
\begin{eqnarray}
D^0(k)\left[D_{++}(q+k)D^0_{++}(k-p) +
D_{--}(q+k)D^0_{--}(k-p)\right]\,. \nonumber
\end{eqnarray}
\noindent Similar transformations of the rest of the integrand
yield
\begin{eqnarray}\label{diagenk6}&&
D^0_{++}(k)D_{+-}(q+k)D^0_{--}(k-p) +
D^0_{--}(k)D_{-+}(q+k)D^0_{++}(k-p) \nonumber\\&&\to -
\frac{1}{2}\left[D^0_{++}(k)D(q+k)D^0_{++}(k-p) +
D^0_{--}(k)D(q+k)D^0_{--}(k-p)\right]\,.
\end{eqnarray}
\noindent Substituting these expressions into Eq.~(\ref{diagenk5})
and using Eq.~(\ref{explprop}) gives
\begin{eqnarray}\label{diagenk61}
\tilde{C}_{\mu\nu\alpha\beta}(p,q) &=&
Q_{\mu\nu}Q_{\alpha\beta}\int\frac{d^4 k}{(2\pi)^3} e^{ik(x-x')}\nonumber\\
&\times&{\rm Re}\left\{ \frac{2\delta(k^2)}{[(k-p)^2 + i0][(q+k)^2
- m^2 + i0]} + \frac{\delta[(q+k)^2 - m^2]}{[k^2 + i0][(k-p)^2 +
i0]}\right\}\,.
\end{eqnarray}
\noindent As was discussed in Sec.~\ref{lr}, the exponent in the
integrand in Eq.~(\ref{diagenk61}) plays the role of an
ultraviolet cutoff, ensuring convergence of the integral at large
$k$'s. On the other hand, the leading contribution (\ref{leading})
is determined by integrating over $k \sim p$ where it is safe to
take the limit $x \to x'.$ Since $(x-x')$ is eventually set equal
to zero, one can further simplify the $k$ integral by using the
dimensional regulator instead of the oscillating exponent. Namely,
introducing the dimensional regularization of the $k$ integral,
one may set $x=x'$ afterwards to obtain
\begin{eqnarray}\label{diagenk7}
\tilde{C}_{\mu\nu\alpha\beta}(p,q) &=&
Q_{\mu\nu}Q_{\alpha\beta}\,\mu^{\epsilon}\,{\rm Re}\int
\frac{d^{4-\epsilon} k}{(2\pi)^3} \left\{
\frac{2\delta(k^2)}{[(k-p)^2 + i0][(q+k)^2 - m^2 + i0]}
\right.\nonumber\\&+& \left. \frac{\delta[(q+k)^2 - m^2]}{[k^2 +
i0][(k-p)^2 + i0]}\right\}\,,
\end{eqnarray}
\noindent where $\mu$ is an arbitrary mass parameter, and
$\epsilon = 4 - d,$ $d$ being the dimensionality of spacetime.

Next, going over to the $\alpha$-representation, the first term in
the integrand may be parameterized as
\begin{eqnarray}&&
\frac{\delta(k^2)}{[(k-p)^2 + i0][(q+k)^2 - m^2 + i0]} =
\frac{\delta(k^2)}{(p^2 - 2kp + i0)(2kq + i0)} \nonumber\\&& =
\frac{1}{2\pi i^2}\iiint_{0}^{\infty}dx dy dz \left(e^{ixk^2} +
e^{-ixk^2}\right)e^{iy(p^2 - 2kp + i0)}e^{iz(2kq + i0)}
\end{eqnarray}
\noindent Substituting this into Eq.~(\ref{diagenk7}), and using
the formulas $$\int d^{4-\epsilon}k~e^{i(a k^2 + 2bk)} = {\rm
sign}(a)\frac{1}{i}\left(\frac{\pi}{|a|}\right)^{2 -
\epsilon/2}\exp\left(\frac{b^2}{ia}\right),$$
$$\int_{0}^{\infty}dx~x^{- \epsilon} e^{ixa} =
i~\Gamma(1-\epsilon)\exp\left({\rm sign
}(a)\frac{\pi\epsilon}{2i}\right)\frac{|a|^\epsilon}{a}\,,$$ one
finds
\begin{eqnarray}
K(p) &\equiv& \mu^{\epsilon}\int \frac{d^{4-\epsilon}
k}{(2\pi)^3}\frac{\delta(k^2)}{[(k-p)^2 + i0][(q+k)^2 - m^2]}
\nonumber\\ &=&
\frac{i\mu^\epsilon}{(2\pi)^4}\iiint_{0}^{\infty}dx dy dz
\left(\frac{\pi}{x}\right)^{2 -
\epsilon/2}e^{iyp^2}\left\{\exp\left[- \frac{i}{x}(y p - z
q)^2\right] - \exp\left[\frac{i}{x}(y p - z q)^2\right]\right\}
\nonumber\\ &=& \frac{\mu^\epsilon\pi^{-\epsilon/2}}{8\pi^2}
\cos\left(\frac{\pi\epsilon}{4}\right) \Gamma\left(1 -
\frac{\epsilon}{2}\right)\iint_{0}^{\infty} dy
dz~e^{iyp^2}\frac{\left|(y p - z q)^2\right|^{\epsilon/2}}{(y p -
z q)^2}\,.\nonumber
\end{eqnarray}\noindent Changing the integration variable $z\to
yz,$ and taking into account that $q^2 = m^2,$ $qp = -p^2/2$
yields
\begin{eqnarray}\label{kintegral}
K(p) &=& \frac{(\mu m)^\epsilon\pi^{-\epsilon/2}}{8\pi^2
m^2}\cos\left(\frac{\pi\epsilon}{4}\right) \Gamma\left(1 -
\frac{\epsilon}{2}\right)\int_{0}^{\infty}dy~e^{i y p^2}
y^{\epsilon - 1}\int_{0}^{\infty}dz \frac{\left|z^2 -
(1+z)\alpha\right|^{\epsilon/2}}{z^2 - (1+z)\alpha} \nonumber\\
&=& \frac{(\pi e^{i\pi})^{-\epsilon/2}}{8\pi^2 m^2}\left(\frac{\mu
m}{- p^2}\right)^{\epsilon}\cos\left(\frac{\pi\epsilon}{4}\right)
\Gamma\left(1 - \frac{\epsilon}{2}\right)
\Gamma(\epsilon)\int_{0}^{\infty}dz~\frac{\left|z^2 -
(1+z)\alpha\right|^{\epsilon/2}}{z^2 - (1+z)\alpha}\ ,
\end{eqnarray}\noindent where $\alpha \equiv - p^2/m^2\,.$
Similar manipulations with the second term in Eq.~(\ref{diagenk7})
give
\begin{eqnarray}\label{int3}
L(p) &\equiv& \mu^{\epsilon}\int \frac{d^{4-\epsilon}
k}{(2\pi)^3}\frac{\delta[(q+k)^2 - m^2]}{[k^2 + i0][(k-p)^2 + i0]}
= - \frac{\mu^{\epsilon}}{(2\pi)^4}\iiint_{0}^{\infty}dx dy dz
\nonumber\\&\times& \int d^{4-\epsilon} k\left(e^{ix(k^2 + 2kq)} +
e^{-ix(k^2+2kq)}\right)e^{iy(p^2 - 2k(q+p) + i0)}e^{iz(-2kq + i0)} \nonumber\\
&=& \frac{i\mu^\epsilon}{(2\pi)^4}\iiint_{0}^{\infty}dx dy dz
\left(\frac{\pi}{x}\right)^{2 - \epsilon/2}e^{iyp^2}
\left\{\exp\left[- \frac{i}{x}((x-z)q - y(q + p))^2\right]
\right.\nonumber\\ &-& \left. \exp\left[\frac{i}{x}((x+z)q + y(q +
p))^2\right]\right\} = \frac{\pi^{-\epsilon/2}}{16\pi^2 m^2
}\left(\frac{\mu}{m}\right)^{\epsilon}\Gamma\left(1 +
\frac{\epsilon}{2}\right)\nonumber\\ &\times&\iint_{0}^{\infty} dy
dz\left\{\frac{e^{i\pi\epsilon/4}}{[(y + z + 1)^2 + yz\alpha]^{1 +
\epsilon/2}} + \frac{e^{-i\pi\epsilon/4}}{[(y + z - 1)^2 +
yz\alpha]^{1 + \epsilon/2}} \right\}\,. \nonumber
\end{eqnarray}\noindent Changing the integration variables $y = ut,$
$z = (1 - u)t$ yields
\begin{eqnarray}\label{int4}
L(p) &=& \frac{\pi^{-\epsilon/2}}{16\pi^2 m^2
}\left(\frac{\mu}{m}\right)^{\epsilon}\Gamma\left(1 +
\frac{\epsilon}{2}\right)\nonumber\\ &\times&\int_{0}^{\infty} dt
\int_{0}^{1}du~t\left\{\frac{e^{i\pi\epsilon/4}}{[(t + 1)^2 + u(1
- u)t^2\alpha]^{1 + \epsilon/2}} + \frac{e^{-i\pi\epsilon/4}}{[(t
- 1)^2 + u(1 - u)t^2\alpha]^{1 + \epsilon/2}} \right\}\,.
\nonumber
\end{eqnarray}\noindent On the other hand, $L(p)$ must be real
because the poles of the functions $D_{++}(k),$ $D_{++}(k - p)$
actually do not contribute. Hence,
\begin{eqnarray}\label{int5}
L(p) &=& \frac{\pi^{-\epsilon/2}}{8\pi^2 m^2
}\left(\frac{\mu}{m}\right)^{\epsilon}\cos\left(\frac{\pi\epsilon}{4}\right)
\Gamma\left(1 + \frac{\epsilon}{2}\right)\int_{0}^{\infty} dt
\int_{0}^{1}\frac{t~d u}{[(t + 1)^2 + u(1 - u)t^2\alpha]^{1 +
\epsilon/2}}\ . \nonumber
\end{eqnarray}\noindent

Let us turn to investigation of singularities of the expressions
obtained when $\epsilon \to 0.$ Evidently, both $K(p)$ and $L(p)$
contain single poles
\begin{eqnarray}\label{div1}
K_{\rm div}(p) &=& \frac{1}{8\pi^2
m^2\epsilon}~\dashint_{0}^{\infty}\frac{dz}{z^2 - (1+z)\alpha} = -
\frac{1}{8\pi^2 m^2\epsilon}\frac{1/\alpha}{\sqrt{1 +
4/\alpha}}\ln\frac{\sqrt{1 + 4/\alpha} + 1}{\sqrt{1 + 4/\alpha} -
1}\ , \\ \label{div2} L_{\rm div}(p) &=& \frac{1}{8\pi^2
m^2\epsilon}~ \int_{0}^{1}\frac{du}{[1 + u(1 - u)\alpha]} =
\frac{1}{4\pi^2 m^2\epsilon}\frac{1/\alpha}{\sqrt{1 +
4/\alpha}}\ln\frac{\sqrt{1 + 4/\alpha} + 1}{\sqrt{1 + 4/\alpha} -
1}\ .
\end{eqnarray}\noindent Note that $L_{\rm div} = - 2 K_{\rm div}.$
Upon substitution into Eq.~(\ref{diagenk7}) the pole terms cancel
each other. Thus, $\tilde{C}_{\mu\nu\alpha\beta}(p,q)$ turns out
to be finite in the limit $\epsilon \to 0.$ Taking into account
also that divergences of the remaining two diagrams (g), (h) in
Fig.~\ref{fig6} are independent of the momentum
transfer,\footnote{The corresponding integrals do not involve
dimensional parameters other than $p,$ and therefore are functions
of $p^2$ only. Hence, on dimensional grounds, both diagrams are
proportional to
$$\left(\frac{c_1}{\epsilon} +
c_2\right)\left(\frac{\mu^2}{-p^2}\right)^{\epsilon/2} =
\frac{c_1}{\epsilon} + c_2 +
\frac{c_1}{2}\ln\left(\frac{\mu^2}{-p^2}\right) + O(\epsilon),$$
where $c_{1,2}$ are some finite constants.} we conclude that the
singular part of the correlation function is completely local.

It is worth of mentioning that although the quantity $K_{\rm
div}(p)$ is non-polynomial with respect to $p^2,$ signifying
non-locality of the corresponding contribution to
$C_{\mu\nu\alpha\beta}(x,x'),$ it is analytic at $p = 0,$ which
implies that its Fourier transform is local to any finite order of
the long-range expansion. Indeed, expansion of $K_{\rm div}(p)$
around $\alpha = 0$ reads
\begin{eqnarray}\label{div11}
K_{\rm div}(p) &=& - \frac{1}{16\pi^2 m^2\epsilon}\left(1 -
\frac{\alpha}{6} + \cdots + \frac{(-1)^n(n!)^2}{(2n + 1)!}\alpha^n
+ \cdots\right)\,.
\end{eqnarray}\noindent
Contribution of such term to $C_{\mu\nu}(x,x')$ is proportional to
\begin{eqnarray}
\int\frac{d^3 \bm{p}}{(2\pi)^3}e^{-i\bm{p}(\bm{x'} - \bm{x}_0
)}\left(1 - \frac{\alpha}{6} + \cdots\right) =
\delta^{(3)}(\bm{x'} - \bm{x}_0) +
\frac{1}{6m^2}\triangle\delta^{(3)}(\bm{x'} - \bm{x}_0) +
\cdots\,. \nonumber
\end{eqnarray}
\noindent The delta function arose here because we neglected
spacial spreading of the wave packet. Otherwise, we would have
obtained
\begin{eqnarray}
\iint\frac{d^3 \bm{q}}{(2\pi)^3}\frac{d^3
\bm{p}}{(2\pi)^3}b^*(\bm{q})b(\bm{q} + \bm{p}) e^{-i\bm{p}(\bm{x'}
- \bm{x}_0 )}\left(1 - \frac{\alpha}{6} + \cdots\right) = 0 \quad
{\rm for} \quad \bm{x'} - \bm{x}_0 \notin W\,, \nonumber
\end{eqnarray}
\noindent as a consequence of the condition (\ref{packet}). In
particular, applying this to the correlation function, we see that
its divergent part does not contribute outside of $W.$ Thus, the
two-point correlation function has a well defined coincidence
limit everywhere except the region of particle localization.

Perhaps, it is worth to stress once more that the issue of
locality of divergences is only a technical aspect of our
considerations. This locality does make the structure of the
long-range expansion transparent and comparatively simple.
However, even if the divergence were nonlocal this would not
present a principal difficulty. A physically sensible definition
of an observable quantity always includes averaging over a finite
spacetime domain, while the singularity of the two-point function,
occurring in the coincidence limit, is integrable (see
Sec.~\ref{lr}). The only problem with the nonlocal divergence
would be impossibility to take the limit of vanishing size of the
spacetime domain [Cf. Eq.~(\ref{c0i0})].

Turning to the calculation of the finite part of $K(p),$ we
subtract the divergence (\ref{div1}) from the right hand side of
Eq.~(\ref{kintegral}), and set $\epsilon = 0$ afterwards:
\begin{eqnarray}\label{int1f}&&
K_{\rm fin}(p)\equiv \lim\limits_{\epsilon\to 0 }\bigl[K(p) -
K_{\rm div}(p)\bigr] \nonumber\\&& = \frac{1}{16\pi^2
m^2}\Biggl\{- \left[i\pi + \ln\pi + \gamma + 2\ln\left(\frac{-
p^2}{\mu m}\right) \right] \dashint_{0}^{\infty}\frac{dz}{z^2 -
(1+z)\alpha} \nonumber\\&&+
~\dashint_{0}^{\infty}dz~\frac{\ln\left|z^2 -
(1+z)\alpha\right|}{z^2 - (1+z)\alpha}\Biggr\}\ ,
\end{eqnarray}\noindent where $\gamma$ is the Euler constant.
The leading contribution is contained in the last integral.
Extracting it with the help of Eq.~(\ref{root}) of the Appendix,
we find
\begin{eqnarray}\label{int1l}
K^{(0)}_{\rm fin}(p) = \frac{1}{64m\sqrt{-p^2}}\,.
\end{eqnarray}\noindent

As to $L(p),$ it does not contain the root singularity. Indeed,
\begin{eqnarray}
L(0) = \frac{\pi^{-\epsilon/2}}{8\pi^2 m^2
}\left(\frac{\mu}{m}\right)^{\epsilon}\cos\left(\frac{\pi\epsilon}{4}\right)
\frac{\Gamma\left(1 + \epsilon/2\right)}{\epsilon(1 +
\epsilon)}\,,\nonumber
\end{eqnarray}\noindent and therefore $L_{\rm fin}(p)\equiv
\lim\limits_{\epsilon\to 0 }[L(p) - L_{\rm div}(p)]$ is finite at
$p=0.$ It is not difficult to verify that $L_{\rm fin}(p)$ is in
fact analytic at $p = 0.$ Thus, substituting Eq.~(\ref{int1l})
into Eqs.~(\ref{diagenk7}), (\ref{diagenk3}), and then into
Eq.~(\ref{c0i0}), using Eq.~(\ref{rootint}), and restoring
ordinary units, we finally arrive at the following expression for
the leading long-range contribution to the correlation function
\begin{eqnarray}\label{final}
\EuScript{C}_{\mu\nu\alpha\beta}^{(0)}(x) =
\delta_{\mu\nu}\delta_{\alpha\beta}\frac{2 m^2 G^2}{c^4 r^2}\ ,
\qquad r = |\bm{x} - \bm{x}_0|\,.
\end{eqnarray}\noindent This result coincides
with that obtained by the author in the framework of the S-matrix
approach \cite{kazakov4}. The nonrelativistic gravitational
potential $\Phi^{\rm g}$ is related to the $00$-component of
metric as $\Phi^{\rm g} = h_{00} c^2/2\,.$ Hence, the root mean
square fluctuation of the Newton potential turns out to be
\begin{eqnarray}\label{rmsp}&&
\sqrt{\left\langle \left(\Delta \Phi^{\rm g}\right)^2
\right\rangle} = \frac{G m}{\sqrt{2}r}\,.
\end{eqnarray}
\noindent Note also that the relative value of the fluctuation is
$1/\sqrt{2}.$ It is interesting to compare this value with that
obtained for vacuum fluctuations. As was shown in
Ref.~\cite{zerbini}, the latter is equal to $\sqrt{2}$ (this is
the square root of the relative variance $\Delta^2_r$ used in
Ref.~\cite{zerbini}).

We can now ask for conditions to be imposed on a system in order
to justify neglecting quantum fluctuations of its gravitational
field. Such a condition can easily be found out by examining
dependence of the $\hbar^0$ contribution on the number of field
producing particles. The structure of $\phi$-lines in the diagram
of Fig.~\ref{fig6}(b) is the same as that in Fig.~\ref{fig2}(a).
Therefore, in the case when the gravitational field is produced by
a $N$-particle body, this diagram is proportional to $m^2 N,$
where $m$ is the mass of a constituent particle. Correspondingly,
the root mean square fluctuation of the potential is proportional
to $m\sqrt{N} = M/\sqrt{N}$ ($M$ is the total mass of the body),
while its relative value, to $1/\sqrt{N}\,.$ We thus arrive to the
condition of macroscopicity of the system, found in
Sec.~\ref{general} in connection with the post-Newtonian loop
contributions. In the present context, it coincides with the
well-known criterion for neglecting quantum gravity fluctuations,
formulated in a somewhat different manner in
Refs.~\cite{tomboulis,callan,zerbini}.

\subsection{Orbit precession in the field of black hole}

Here the results obtained in the preceding sections will be
applied to the investigation of dynamics of a classical particle
in the gravitational field of a black hole with mass $M.$ The
particle will be taken {\it testing}, i.e., it will be assumed
sufficiently light to neglect its contribution to the
gravitational field, and sufficiently small compared with $r_{\rm
g} = 2 G M/c^2$ to allow considering it as a pointlike object. The
latter assumption, in particular, justifies the use of the
expression (\ref{final}) for the correlation function, obtained in
the coincidence limit.

In order to find equations of motion of the particle, we have to
calculate its effective action. The action functional for a
pointlike particle has the form
\begin{eqnarray}&&
S_{\rm t} = - m_{\rm t}{\displaystyle\int}
\sqrt{g_{\mu\nu}dx^{\mu}dx^{\nu}} = - m_{\rm t}
{\displaystyle\int}d\tau\sqrt{1 + h_{\mu\nu}\dot x^{\mu} \dot
x^{\nu}}\ , \qquad \dot x^{\mu} \equiv \frac{dx^\mu}{d\tau}
\nonumber
\end{eqnarray}
\noindent where $d\tau$ is the particle proper time defined with
respect to the flat metric, $d\tau^2 = \eta_{\mu\nu}d x^{\mu} d
x^{\nu}\,.$ Inserting this expression into the generating
functional integral of Green functions, one can write, in view of
the assumed smallness of the test particle contribution,
\begin{eqnarray}&&
\int\EuScript{D}\Phi_-\int\EuScript{D}\Phi_+\exp\left\{i(S +
S_{\rm t} + j\phi) \right\} \nonumber\\&& = \left.Z_{\rm
SK}\right|_{\genfrac{}{}{0pt}{}{J\setminus j = 0}{K = 0}} +
\int\EuScript{D}\Phi_-\int\EuScript{D}\Phi_+\,i S_{\rm
t}\,\exp\left\{i(S + j\phi)\right\} \nonumber\\&& = \left.Z_{\rm
SK}\left(1 + i\langle {\rm in}|S_{\rm t}|{\rm
in}\rangle\right)\right|_{\genfrac{}{}{0pt}{}{J\setminus j = 0}{K
= 0}}\,.
\end{eqnarray}
\noindent The generating functional of connected Green functions
takes the form $$W = - i \ln\left\{ Z_{\rm SK}\left(1 + i\langle
{\rm in}|S_{\rm t}|{\rm in}\rangle\right)\right\} = W_{\rm SK} +
\langle {\rm in}|S_{\rm t}|{\rm in}\rangle\,.$$ Taking its
Legendre transform, we see that the effective particle action is
just
$$\Gamma_{\rm t} = \langle {\rm in}|S_{\rm t}|{\rm in}\rangle.$$
Expanding $S_{\rm t}$ in powers of $h_{\mu\nu},$ the right hand
side of this equation can be evaluated in the first post-Newtonian
approximation as
\begin{eqnarray}\label{tpaction}
\Gamma_{\rm t} &=& - m_{\rm t}{\displaystyle\int} d \tau \left\{1
+ \frac{1}{2}\langle h_{\mu\nu}\rangle\dot x^{\mu}\dot
x^{\nu}-\frac{1}{8}\left[\langle h_{\mu\nu}\rangle \langle
h_{\alpha\beta}\rangle +
\EuScript{C}^{(0)}_{\mu\nu\alpha\beta}\right]\dot x^{\mu}\dot
x^{\nu}\dot x^{\alpha}\dot x^{\beta} \right\}
\nonumber\\
&=& - m_{\rm t}{\displaystyle\int} \sqrt{\left(\eta_{\mu\nu} +
\langle h_{\mu\nu}\right\rangle) d x^{\mu}d x^{\nu} -
\frac{1}{4}\EuScript{C}_{0000}^{(0)}(d x^0)^2}\ .
\end{eqnarray}
\noindent Substituting Eqs.~(\ref{eff}), (\ref{final}) for the
mean gravitational field and its fluctuation, we see from
Eq.~(\ref{tpaction}) that the test particle motion in the
fluctuating field of a black hole is effectively the same as in a
non-fluctuating gravitational field described by the following
spacetime metric
\begin{eqnarray}\label{eff1}&&
d\bar{s}^2 \equiv \bar{g}^{\rm eff}_{\mu\nu} d x^{\mu} d x^{\nu} =
\left(1 - \frac{r_g}{r} - \frac{r^2_g}{8 r^2}\right) d t^2 -
\left(1 + \frac{r_g}{r} - \frac{7 r^2_g}{4 r^2}\right) d r^2
\nonumber\\&& - r^2 \left(1 + \frac{r_g}{r} + \frac{7 r^2_g}{4
r^2}\right) (d\theta^2 + \sin^2\theta\ d\varphi^2),
\end{eqnarray}

Now, it is not difficult to calculate the secular precession of
the particle's orbit. Let the test particle with move in the
equatorial plane ($\theta = \pi/2$) around black hole. Then we
have the following Hamilton-Jacobi equation for the action
$\Gamma_{\rm t}$
$$\bar{g}^{\mu\nu}_{\rm eff}\frac{\partial \Gamma_{\rm t}}{\partial x^{\mu}}
\frac{\partial \Gamma_{\rm t}}{\partial x^{\nu}} - m_{\rm t}^2 =
0,$$ where $\bar{g}^{\mu\nu}_{\rm eff}$ is the reciprocal of
$\bar{g}_{\mu\nu}^{\rm eff}.$ A simple calculation gives, to the
leading order,
\begin{eqnarray}\label{actest}
S_{\rm b} &=& - E t + L\varphi \nonumber\\ &+& \int dr
\left[\left(E'^2 + 2 m_{\rm t} E'\right) +
\frac{r_g}{r}\left(m_{\rm t}^2 + 4 m_{\rm t} E'\right) -
\frac{1}{r^2}\left(L^2 - \frac{17}{8} r_g^2 m_{\rm t}^2
\right)\right]^{1/2},
\end{eqnarray} \noindent where $E, L$ are the energy and angular
momentum of the particle, respectively, and $E' = E - m_{\rm t}$
its non-relativistic energy. The first two terms in the integrand
in Eq.~(\ref{actest}) coincide with the corresponding terms of
classical theory, while the third does not, leading to the angular
shift of the orbit
\begin{eqnarray}\label{prec} \delta \varphi =
\frac{17\pi G M}{2 c^2 a (1 - e^2)}
\end{eqnarray} \noindent
per period ($a$ and $e$ are the major semiaxis and the
eccentricity of the orbit, respectively).

\section{Concluding remarks}\label{conclud}

In this Chapter, we have determined the long-range behavior of the
gravitational fields produced by quantized matter in the first
post-Newtonian approximation. From the point of view of
establishing the correct correspondence between classical and
quantum theories of gravitation, Eq.~(\ref{mainnew}) demonstrating
violation of general covariance by the $\hbar^0$-order loop
contributions, and Eq.~(\ref{final}) describing
$\hbar^0$-fluctuations of the spacetime metric turned out to be of
special importance, and led us independently to a macroscopic
formulation of the correspondence principle. In turn, this
formulation endows $\hbar^0$ loop contributions with direct {\it
physical meaning as describing deviations of the spacetime metric
from classical solutions of the Einstein equations in the case of
finite number of elementary particles constituting the gravitating
body.}

In connection with these results, it is worth also to make the
following general comments.

As we have seen, an essentially quantum character of elementary
particle interactions makes classical consideration inapplicable
to systems whose dynamics is governed by interactions of a finite
number of constituent particles. On the other hand, there is a
deep-rooted belief in the literature that the quantum field
description of interacting remote systems, each of which consists
of many particles, is equivalent to that in which these systems
are replaced by elementary particles with masses and charges equal
to the total masses and charges of the systems. In other words,
without calling it into question, the familiar notion of a point
particle is carried over from classical mechanics to quantum field
theory. This point of view is adhered, for instance, in the
classic paper by Iwasaki \cite{iwasaki} where it is applied to the
solar system to calculate the secular precession of Mercury's
orbit, considered as a ``Lamb shift''. The Sun and Mercury are
regarded in Ref.~\cite{iwasaki} as scalar particles. As we saw in
Sec.~\ref{calcul}, under this assumption the root mean square
fluctuation of the solar gravitational potential would be $71\%$
of its mean value. Fortunately, such fluctuations are not observed
in reality. This is because the Sun is composed of a huge number
of elementary particles each of which contributes to the total
gravitational field. As we have seen, the relative quantum
fluctuation turns out to be suppressed in this case by the factor
$1/\sqrt{N} \sim \sqrt{m_{\rm proton}/M_{\odot}} \approx
10^{-28}\,.$

Another example of attempts to recover the nonlinearity of a
classical theory through the radiative corrections can be found in
Ref.~\cite{donoghue2}. The authors of \cite{donoghue2} claim that
the electromagnetic corrections of the order $e^2$ to the
classical Reissner-Nordstr\"{o}m solution are reproduced by the
diagram of Fig.~\ref{fig2}(a) in which the internal wavy lines
correspond to the virtual photons. However, as we have shown, it
is meaningless to try to establish the correspondence between
classical and quantum theories in terms of {\it elementary}
particles, because quantum fluctuations of the electromagnetic and
gravitational fields produced by such particles are of the order
of the fields themselves. On the other hand, because of its
inappropriate dependence on the number of particles, the diagram
\ref{fig2}(a) fails to reproduce the classical physics in the
macroscopic limit. This can be shown using the same argument as in
the case of purely gravitational interaction. Namely, given a body
with the total electric charge $Q,$ consisting of $N = Q/q$
particles with charge $q,$ the contribution of the diagram
\ref{fig2}(a) is proportional to $N\cdot q^2 = Q^2/N\,$ turning
into zero in the macroscopic limit. The relevant contribution
correctly reproducing the $e^2$-correction to the
Reissner-Nordstr\"{o}m solution is given instead by the tree
diagrams like that in Fig.~\ref{fig1}(b) in which internal wavy
lines correspond to virtual photons.

Finally, we mention that investigation of quantum fluctuations of
the Coulomb potential similar to that carried out in
Sec.~\ref{qflucs} shows that the macroscopic formulation of the
correspondence principle given above extends also to the case of
electromagnetic interaction \cite{kazakov4,kazakov5}.

\begin{appendix}

\section{Root singularities of Feynman integrals}

The integrals
\begin{eqnarray}&&
J_{nm} \equiv \int_{0}^{\infty}\int_{0}^{\infty} \frac{du dt}{(1 +
t + u)^n (1 + \alpha t u)^m}, \nonumber
\end{eqnarray}
encountered in Sec.~\ref{general}, can be evaluated as follows.
Consider the auxiliary quantity
\begin{eqnarray}&&
J(A,B) = \int_{0}^{\infty}\int_{0}^{\infty} \frac{du dt}{(A + t +
u) (B + \alpha t u)}, \nonumber
\end{eqnarray}
\noindent where $A,B>0$ are some numbers eventually set equal to
1. Performing an elementary integration over $u,$ we get
\begin{eqnarray}&&
J(A,B) = \int_{0}^{\infty} dt ~\frac{\ln B - \ln \{\alpha t (A +
t)\}}{B - \alpha t (A + t)}. \nonumber
\end{eqnarray}

\begin{figure}
\includegraphics{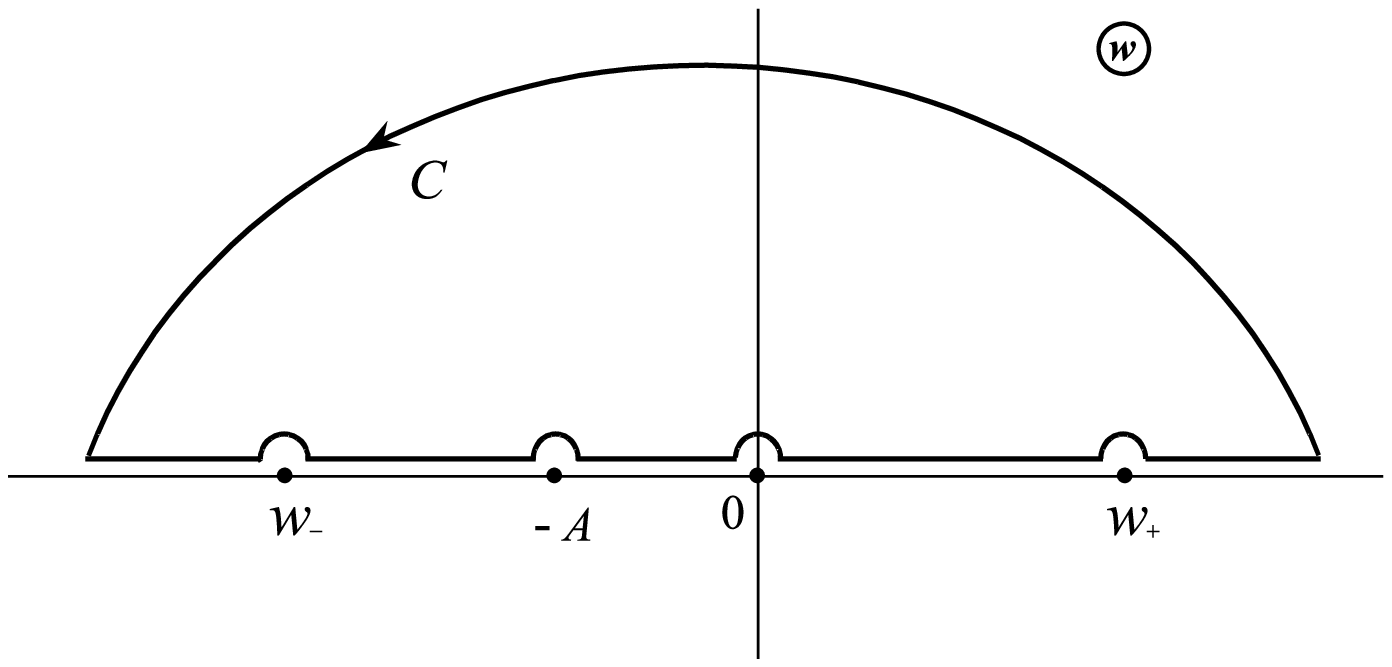}
\caption{Contour of integration in Eq.~(\ref{cint1}).}\label{fig7}
\end{figure}

Now consider the integral
\begin{eqnarray}&&\label{cint1}
\tilde{J}(A,B) = \int_{C} d w f(w,A,B), ~~f(w,A,B) = \frac{\ln B -
\ln \{\alpha w (A + w)\}}{B - \alpha w (A + w)},
\end{eqnarray}
taken over the contour $C$ shown in Fig.~\ref{fig7}.
$\tilde{J}(A,B)$ is zero identically. On the other hand,
\begin{eqnarray}&&
\tilde{J}(A,B) = \int^{-A}_{- \infty} dz ~\frac{\ln B - \ln
\{\alpha z (A + z)\}}{B - \alpha z (A + z)} + \int_{-A}^{0} dz
~\frac{\ln B - \ln \{ - \alpha z (A + z)\} + i\pi}{B - \alpha z (A
+ z)} \nonumber\\&& + ~\dashint_{0}^{+ \infty} dz ~\frac{\ln B -
\ln \{\alpha z (A + z)\} + 2 i\pi}{B - \alpha z (A + z)} - i\pi
\sum\limits_{w_{+},w_{-}} {\rm Res} f(w,A,B), \nonumber
\end{eqnarray}
\noindent $w_{\pm}$ denoting the poles of the function $f(w,A,B),$
$$w_{\pm} = - \frac{A}{2} \pm \sqrt{\frac{B}{\alpha} + \frac{A^2}{4}}.$$
Change $z \to - A - z$ in the first integral. A simple calculation
then gives
\begin{eqnarray}&&\label{int4new}
J(A,B) = \frac{\pi^2}{2\sqrt{\alpha}} B^{-1/2} \left(1 +
\frac{\alpha A^2}{4 B}\right)^{-1/2} - \frac{1}{2}\int_{0}^{A} dt
~\frac{\ln B - \ln \{\alpha t (A - t)\}}{B + \alpha t (A - t)}.
\end{eqnarray}

The roots are contained entirely in the first term on the right of
Eq.~(\ref{int4new}). The integrals $J_{nm}$ are found by repeated
differentiation of Eq.~(\ref{int4new}) with respect to $A,B$.
Expanding $(1 + \alpha A^2/4 B)^{-1/2}$ in powers of $\alpha,$ we
find the leading terms needed in Sec.~\ref{general}
\begin{eqnarray}\label{roots}
J^{{\rm root}}_{12} &=& \frac{\pi^2}{4\sqrt{\alpha}}, ~~J^{{\rm
root}}_{13} = \frac{3\pi^2}{16\sqrt{\alpha}}, \nonumber\\ J^{{\rm
root}}_{31} &=& - \frac{\pi^2}{16}\sqrt{\alpha}, ~~J^{{\rm
root}}_{32} = - \frac{3\pi^2}{32}\sqrt{\alpha}, ~~J^{{\rm
root}}_{33} = - \frac{15\pi^2}{128}\sqrt{\alpha}.
\end{eqnarray}

\begin{figure}
\includegraphics{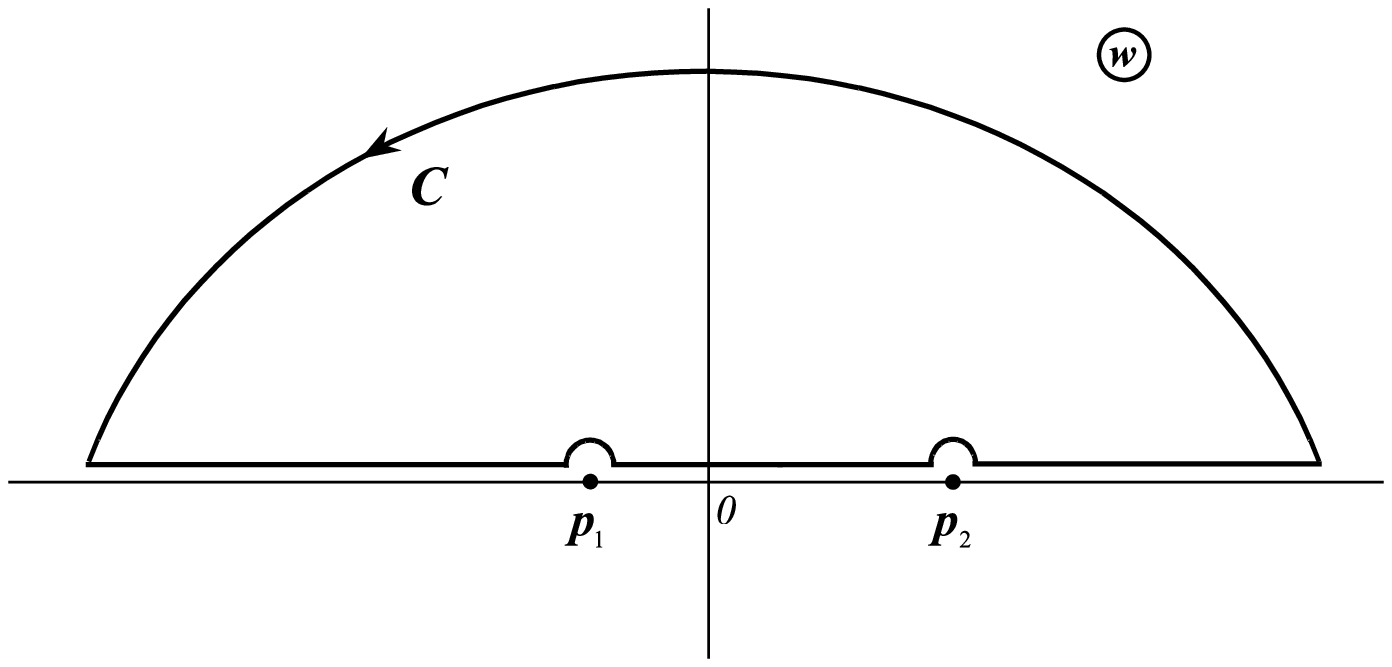}\caption{Contour of integration in
Eq.~(\ref{cint2}).}\label{fig8}
\end{figure}

Next, in the course of evaluation of the integral $K(p)$ we
encountered the integral
\begin{eqnarray}
A \equiv \dashint_{0}^{\infty}dz~\frac{\ln\left|z^2 - (1 + z
)\alpha\right|}{z^2 - (1 + z )\alpha} &=&
\dashint_{0}^{\infty}dz~\frac{\ln\left|(z - p_1)(z -
p_2)\right|}{(z - p_1)(z - p_2)}\,, \nonumber\\ p_{1,2} &=&
\frac{\alpha \pm\sqrt{\alpha^2 + 4\alpha}}{2}\,. \nonumber
\end{eqnarray}
\noindent To evaluate this integral, let us consider an auxiliary
integral
\begin{eqnarray}\label{cint2}
\bar{A} = \int_{C}d w~\frac{\ln[(w - p_1)(w - p_2)]}{(w - p_1)(w -
p_2)}\ ,
\end{eqnarray}
\noindent taken over the contour $C$ in the complex $w$ plane,
shown in Fig.~\ref{fig8}. $\bar{A}$ is zero identically. On the
other hand, for sufficiently small positive $\sigma,$ one has
\begin{eqnarray}&&
\bar{A} = \int_{-\infty}^{p_1 - \sigma}dz~\frac{\ln[(z - p_1)(z -
p_2)]}{(z - p_1)(z - p_2)} - i\pi \frac{\ln(p_2 - p_1)}{p_1 - p_2}
- \frac{i\pi\ln\sigma + \pi^2/2}{p_1 - p_2} \nonumber\\&& +
\int_{p_1 + \sigma}^{p_2 - \sigma}dz~\frac{\ln[(z - p_1)(p_2 - z)]
- i\pi}{(z - p_1)(z - p_2)} - i\pi \frac{\ln(p_2 - p_1) -
i\pi}{p_2 - p_1} - \frac{i\pi\ln\sigma + \pi^2/2}{p_2 - p_1}
\nonumber\\&& + \int_{p_2 + \sigma}^{\infty}dz~\frac{\ln[(z -
p_1)(z - p_2)] - 2i\pi}{(z - p_1)(z - p_2)}\,. \nonumber
\end{eqnarray}
\noindent Changing the integration variable $z \to (p_1 + p_2) -
z$ in the first integral, and rearranging yields in the limit
$\sigma \to 0$
\begin{eqnarray}\label{eroot}
A = \frac{1}{2}\int_{0}^{p_1 + p_2} dz~\frac{\ln[(p_2 - z)(z - p_1
)]}{(z - p_1)(z - p_2)} + \frac{\pi^2/2}{p_2 - p_1}\,.
\end{eqnarray}
\noindent The root singularity is contained in the second term,
because
$$\int_{0}^{p_1 + p_2} dz~\frac{\ln[(p_2 - z)(z - p_1 )]}{(z -
p_1)(z - p_2)} \to - \ln\alpha \quad {\rm for} \quad \alpha\to
0.$$ Thus,
\begin{eqnarray}\label{root}
A^{\rm root} = \frac{\pi^2}{4\sqrt{\alpha}}\,.
\end{eqnarray}
\noindent

\end{appendix}

\pagebreak

\pagebreak

\end{document}